%% file: RR-9491.tex
\documentclass[twoside]{article}
\usepackage[a4paper]{geometry}
\usepackage[utf8]{inputenc}
\usepackage[T1]{fontenc} 
\usepackage{RR}
\usepackage{hyperref}
\usepackage{graphicx}
\usepackage{subfig}
\usepackage{caption}
\usepackage{amsfonts,latexsym,amssymb,amscd,epsf}
\usepackage{amsthm}
\usepackage[tbtags]{amsmath}
\usepackage{mathtools}

\usepackage{algorithm}
\usepackage{algpseudocode}
\algnewcommand{\LineComment}[1]{\Statex \(\triangleright\) \small #1}
\algnewcommand{\Input}[1]{\Statex \textbf{input:} #1}
\algnewcommand{\Output}[1]{\Statex \textbf{output:} #1}
\usepackage{booktabs,multirow}

\usepackage{array}
\usepackage{makecell}

\usepackage{dsfont}
\usepackage{bm}				
\DeclareMathAlphabet{\mymathbb}{U}{BOONDOX-ds}{m}{n}
\DeclareMathAlphabet\mathbfcal{OMS}{cmsy}{b}{n}
\newcommand*{\myfont}{\fontfamily{qzc}\selectfont}
\newcommand{\set}[1]{\text{\large\myfont{#1}\normalsize}}

\newcommand{\op}[1]{\mathbfcal{#1}}
\newcommand{\numberset}{\mathbb}
\newcommand{\ten}[1]{\mathbf{#1}}

\newcommand{\mat}{\text{mat}}

\newcommand{\I}{\mathbb{I}}

\newcommand{\R}{\numberset{R}}

\newcommand{\N}{\numberset{N}}

\newcommand{\coreTT}[1]{\underline{\ten{#1}}}
\newcommand{\matcoreTT}[1]{\underline{{#1}}}
\usepackage{tensor}
 
\newcommand{\dotprod}[2]{{\langle #1, #2\rangle}}
\newcommand{\norm}[1]{\Vert{#1} \Vert}

\renewcommand{\vec}[1]{#1}
\renewcommand{\mat}[1]{#1}
\newcommand{\bigO}{\mathcal{O}}
\newcommand{\ignore}[1]{}

\newcommand{\hypref}[2]{\hyperref[#2]{#1 \ref*{#2}}}

\newcommand{\MI}[1]{\ignore{\color{blue}[MI: #1]}} 

\usepackage{etoolbox} 

\DeclareMathAlphabet{\mathdutchcal}{U}{dutchcal}{m}{n}
\SetMathAlphabet{\mathdutchcal}{bold}{U}{dutchcal}{b}{n}
 \newtheorem{theorem}{Theorem}[section]

 \newtheorem{remark}[theorem]{Remark}

\usepackage[style=numeric, backend=bibtex, firstinits, maxbibnames=50,sorting=none, url=false,eprint=false]{biblatex}
\bibliography{references}

\RRNo{9491}
\RRdate{November 2022}
 \RRversion{2}
 \RRdater{January 2024}
\RRauthor{
  Olivier Coulaud\thanks{Inria, Inria centre at the University of Bordeaux}, 
  \and
  Luc Giraud\footnotemark[1]
  \and
  Martina Iannacito\thanks{Group Science, Technology and Engineering, KU Leuven - Kulak, Departement of Electrical Engineering, KU Leuven, Belgium}, 
}

\authorhead{Coulaud et al.}
\RRtitle{À propos de schémas d'orthogonalisation dans le format Tensor Train}
\RRetitle{On some orthogonalization schemes in Tensor Train format}
\titlehead{On some orthogonalization schemes in TT-format}
\RRresume{Dans le contexte de l'espace tensoriel, nous considérons les noyaux d'orthogonalisation pour générer une base orthogonale d'un sous-espace tensoriel à partir d'un ensemble de tenseurs linéairement indépendants. En particulier, nous étudions numériquement la perte d'orthogonalité de six méthodes d'orthogonalisation, à savoir les méthodes de Gram-Schmidt classique et modi\-fiée avec (CGS2, MGS2) et sans (CGS, MGS) réorthogonalisation, l'approche de Gram et la transformation de Householder. Pour lutter contre la malédiction de la dimensionnalité, nous représentons les tenseurs avec le formalisme Train de tenseurs (TT), et nous introduisons des étapes de recompression dans le schéma de l'algorithme standard par la méthode TT-rounding à une précision prescrite. Après avoir décrit la structure et les propriétés de l'algorithme, nous illustrons numériquement que les limites théoriques de la perte d'orthogonalité dans le calcul matriciel classique sont maintenues, l'arrondi unitaire étant remplacé par la précision de l'arrondi TT. L'analyse pour chaque noyau d'orthogonalisation de l'exigence de mémoire et de la complexi\-té de calcul en termes d'arrondi TT, qui se trouve être l'opération la plus coûteuse en termes de calcul, complète l'étude.}

\RRabstract{In the framework of tensor spaces, we consider orthogonalization kernels to generate an orthogonal basis of a tensor subspace from a set of linearly independent tensors. In particular, we experimentally study the loss of orthogonality of six orthogonalization methods, namely Classical and Modified Gram-Schmidt with (CGS2, MGS2) and without (CGS, MGS) re-orthogonalization, the Gram approach, and the Householder transformation. To overcome the curse of dimensionality, we represent tensors with a low-rank approximation using the Tensor Train (TT) formalism. In addition, we introduce recompression steps in the standard algorithm outline through the TT-rounding method at a prescribed accuracy. After describing the structure and properties of the algorithms, we illustrate their loss of orthogonality with numerical experiments. The theoretical bounds from the classical matrix computation round-off analysis, obtained over several decades, seem to be maintained, with the unit round-off replaced by the TT-rounding accuracy. The computational analysis for each orthogonalization kernel in terms of the memory requirements and the computational complexity measured as a function of the number of TT-rounding, which happens to be the most computationally expensive operation, completes the study.}
\RRmotcle{Gram-Schmidt classique, Gram-Schmidt modifiée, transformation de Householder, perte d'orthogonalité, Tensor Train format}
\RRkeyword{Classical Gram-Schmidt, Modified Gram-Schmidt, Householder transformation, loss of orthogonality, Tensor Train format}
\RRprojet{Concace}
\RCBordeaux 

\begin{document}
\makeRR   
\let\thefootnote\relax\footnotetext{
Distributed under a \href{https://creativecommons.org/licenses/by/4.0/}{Creative Commons Attribution 4.0} International License
}

\tableofcontents
\newpage
\input{core.tex}

\printbibliography

\end{document}

%% file: core.tex
\section{Introduction}\label{sec0}
	 The solution of linear problems is at the heart of many large-scale simulations in academic or industrial applications. Many numerical linear algebra algorithms rely on an orthonormal basis of the space in which the solution is sought; this is particularly the case in GMRES, one of the most popular Krylov subspace methods for solving linear systems, or all variants of the Arnoldi algorithms for computing eigenpairs~\cite{Saad2003, Saad2011EV}. The orthonormal basis is built from a set of vectors that are explicitly orthonormalized by an orthogonalization procedure. Various orthogonalization algorithms have been proposed to perform this task over the years. Additionally, they allow for the computation of the matrix \textsc{QR} factorization. If the input consists of $m$ vectors of $\R^{n}$ organized as the columns of a matrix $A\in\R^{n\times m}$, then the orthogonalization schemes can factorize $A$ into the product of an orthogonal matrix $Q\in\R^{n\times m}$ and an upper triangular one $R\in\R^{m\times m}$. Among the most widely used numerical algorithms, we consider the Classical Gram-Schmidt (CGS)~\cite{Gram1883, Schmidt1907}, the Modified Gram-Schmidt (MGS)~\cite{Gram1883, Schmidt1907}, their variants with re-orthogonalization, named CGS2 and MGS2~\cite{Abdelmalek1971, Daniel1976, parlett1980}, the Gram approach~\cite{Stathopoulos2002} and the Householder transformations~\cite{Householder1958}. CGS and MGS are algorithms that implement the Gram-Schmidt method. The fundamental idea is to sequentially remove the projection of an input vector along the previously computed orthonormal vectors and eventually normalize it. The CGS2 and MGS2 procedures aim to improve the quality of the CGS and MGS basis vectors by orthogonalizing them once more in the same way as the basis computed with CGS and the MGS, respectively. The Gram method calculates the orthogonal basis by utilizing the Cholesky factorization of the Gram matrix, which is defined by the inner products of input vectors. The Householder transformation relies on orthogonal reflections constructed from the input vectors and used to reflect the canonical basis.
	\par A crucial aspect of finite precision calculation for orthogonalization algorithms is the \emph{loss of orthogonality} in the computed basis due to computational rounding errors. 
	This issue has been extensively studied over the years, resulting in numerous findings. The research articles present many theoretical results that relate the loss of orthogonality to the linear dependency of the input vectors. The authors of~\cite{Bjorck1967,Giraud2005} establish theoretical bounds for CGS and MGS loss of orthogonality, showing that the basis produced by MGS is better in terms of orthogonality than the CGS one. In~\cite{Giraud2005} for CGS2 and MGS2, it is confirmed that this re-orthogonalization effectively improves the orthogonality of the computed basis. Bounds for the loss of orthogonality of the Householder transformation and the Gram method are proven in~\cite{Wilkinson1971} and~\cite{Stathopoulos2002}, respectively. Collectively, these theoretical results are several decades old. From Wilkinson's oldest paper in 1965 for the Householder transformation to the most recent method presented in 2006 by Barlow et al. for the CGS2 scheme.
	 
	\par All of the cited algorithms translate naturally into the tensor world. Starting from a set of $m$ tensors of $\R^{n_1\times \dots\times n_d}$, an orthogonal basis for the relative subspace of dimension $m$ of $\R^{n_1\times \dots\times n_d}$ is produced. These kernels are used in iterative methods to solve linear systems structured with the tensor product or in generalization of the least-squares problem to the tensor space. These orthogonalization schemes can work with dense tensors, but they are affected by the ``curse of dimensionality'', i.e., their storage and operation costs grow exponentially with the order of the tensor. Therefore, it is necessary to represent the tensor in a compressed format. In this work, we generalize the six orthogonalization kernels previously mentioned to tensors, using the Tensor Train (TT) formalism~\cite{Oseledets2011TT, Oseledets2009}. These kernels in TT-format can be used, for example, in TT-algorithms such as TT-GMRES~\cite{Dolgov2013}. 
	However, the operation sequences between tensors in TT-formats reduce the benefit of this compressed representation. Therefore, we introduce additional compression steps by the \texttt{TT-round} function~\cite{Oseledets2011TT} in the orthogonalization schemes, knowing that they affect the orthogonality quality of the basis. The aim of this work is to describe the orthogonalization kernels generalized to the tensor with the TT-format and to experimentally investigate the loss of orthogonality of the computed basis. These numerical results in TT-format show a similarity with the theoretical results of classical numerical matrix computation. 	 	
	\par The rest of the paper is organized as follows. In Section~\ref{sec1}, we introduce the notation and recall the most important properties of the TT-format. Section~\ref{sec2} begins with a description of the six orthogonalization schemes extended to the tensor context by the TT-formalism. We also address the complexity in terms of the number of \texttt{TT-round} applications, which is the most computationally expensive operation. In Section~\ref{sec2:s4} we recall briefly the known theoretical bounds related to the loss of orthogonality of these schemes in classical matrix computation. The theoretical results are linked to the numerical experiments, collected in Section~\ref{sec3}, of the same orthogonalization schemes extended with the TT-format. The similarities between the classical orthogonalization kernels and their TT-versions are summarized in Section~\ref{sec4}.

	\section{Notation and TT-format}\label{sec1}
	
	To enhance readability, we utilize the following notations for the various mathematical objects described. Small Latin letters represent scalars and vectors (e.g., $a$), with the context clarifying the object's nature. Capital Latin letters denote matrices (e.g., $A$), while bold small Latin letters denote tensors (e.g., $\ten{a}$). Calligraphic capital letters represent sets (e.g., $\set{A}$). We use the `Matlab notation' to indicate all the indices along a mode with a colon (`$:$'). For example, if we are given a matrix $A\in\R^{m\times n}$, then $A(:, i)$ represents the $i$-th column of $A$. The tensor product is denoted by $\otimes$, while the {Euclidean} inner product is denoted by $\dotprod{\cdot}{\cdot}$ for both vectors and tensors. We use $||\cdot||$ to denote the Euclidean norm for vectors and the Frobenious norm for matrices and tensors. The condition number of a matrix $A\in\R^{n\times n}$ is denoted by $\kappa(A) = \big\|A \big\|\norm{A^{-1}}$.
	\par Let $\ten{x}$ be a $d$-order tensor in $\R^{n_1\times \dots \times n_d}$ and $n_k$ the {dimension} of mode $k$ for every $k\in\{1, \dots, d\}$. Storing the full tensor $\ten{x}\in\R^{n_1\times \dots \times n_d}$ has a memory cost of $\mathcal{O}(n^{d})$ with $n=~\max_{i\in\{1,\dots, d\}}\{n_i\}$. Therefore, various compression techniques have been proposed over the years to reduce the memory consumption~\cite{Lathauwer2000a, Grasedyck2009, Oseledets2011}. For the purpose of this work the most suitable tensor representation is the \emph{Tensor Train} (TT) format~\cite{Oseledets2011}. The main concept of TT is to represent a $d$-order tensor as the contraction of $d$ $3$-order tensors. This contraction is a generalization of the matrix-vector product to tensors.

	The Tensor Train representation of $\ten{x}\in\R^{n_1\times \dots \times n_d}$ is
	\[
	\ten{x} = \coreTT{x}_1\coreTT{x}_2\cdots\coreTT{x}_d ,
	\]
	where $\coreTT{x}_k\in\R^{r_{k-1}\times n_k\times r_k}$ is called $k$-th \emph{TT-core} for $k\in\{1,\dots, d\}$, with $r_0 = r_d = 1$. Note that $\coreTT{x}_1\in\R^{r_0\times n_1\times r_1}$ and $\coreTT{x}_d\in\R^{r_{d-1}\times n_d\times r_d}$ reduce essentially to matrices, but for consistency in notation, we represent them as tensors. The $k$-th TT-core of a tensor is denoted by the same bold letter underlined with a subscript $k$. The value $r_k$ is called \emph{$k$-th TT-rank}. 

		
	Given an index $i_k$, we denote the $i_k$-th matrix slice of $\coreTT{x}_k$ with respect to mode $2$ by $\matcoreTT{X}_k(i_k)$, i.e., $\matcoreTT{X}_k(i_k) = \coreTT{x}_k(:, i_k, :)$. Each element of the TT-tensor $\ten{x}$ can be expressed as the product of $d$ matrices, i.e.,
	\[
	\ten{x}(i_1,\dots, i_d) = \matcoreTT{X}_1(i_1)\cdots\matcoreTT{X}_{d}(i_d)
	\]
	with $\matcoreTT{X}_k(i_k)\in\R^{r_{k-1}\times r_k}$ for every $i_k\in\{1,\dots, n_k\}$ and $k\in\{2, \dots, d-1\}$, while $\matcoreTT{X}_{1}(i_1)\in\R^{1\times r_1}$ and $\matcoreTT{X}_d(i_d)\in\R^{r_{d-1}\times 1}$. It is important to note that $\matcoreTT{X}_{1}(i_1)$ and $\matcoreTT{X}_d(i_d)$ are actually vectors, but for the sake of consistency, they are written as matrices with a single row or column.
	
	Storing a tensor in TT-format requires $\mathcal{O}(dnr^2)$ units of memory, where $n = \max_{i\in\{1,\dots, d\}}\{n_i\}$ and $r = \max_{i\in\{1,\dots, d\}}\{r_i\}$. The memory footprint grows linearly with the tensor order and quadratically with the maximal TT-rank. Therefore, knowing the maximal TT-rank is usually sufficient to estimate the TT-compression benefit. However, for greater accuracy, we introduce the compression ratio measure.
	If $\ten{x}\in\R^{n_1\times\dots\times n_d}$ is a tensor in TT-format, then the compression ratio is the ratio between the storage cost of $\ten{x}$ in TT-format and the storage cost in dense format, i.e.,
	\begin{equation}
		\label{eq:comp_ratio}
		\dfrac{\sum_{i = 1}^{d}r_{i-1}n_ir_i}{\prod_{j=1}^{d}n_j}
	\end{equation}
	where $r_i$ is the $i$-th TT-rank of $\ten{x}$.
	As demonstrated by the compression ratio, to significantly benefit from this formalism, the TT-ranks $r_i$ must remain bounded and small. However, some operations among tensors in TT-format, such as algebraic addition, can increase the TT-ranks. For instance, given two TT-tensors $\ten{x}$ and $\ten{y}$ with $k$-th TT-rank $r_k$ and $s_k$ respectively, then the $k$-th TT-rank of $\ten{x} + \ten{y}$ is equal to $r_k+s_k$, see~\cite{Gel2017TheTF}. 
	To address the issue of the TT-rank growth, a rounding algorithm to reduce it was proposed in~\cite{Oseledets2011}. The \texttt{TT-round} algorithm takes a TT-vector $\ten{x}$ and a relative accuracy $\delta$ as inputs and returns a TT-tensor $\ten{\tilde{x}}$, that is at a relative distance $\delta$ from $\ten{x}$, i.e., $|| \ten{x}-\ten{\tilde{x}}|| \le \delta ||\ten{x}||$. The TT-round function is fully described in~\cite{Oseledets2011TT}. For large-scale tensors, a randomized version of the TT-round function is described in~\cite{AlDaas2023, Che2019}.
	
	To evaluate the benefit of the \texttt{TT-round}, we introduce the \emph{compression gain}, which is the ratio of the compression ratios, written as
	\begin{equation}
		\label{eq:gain_ratio}
			\dfrac{\sum_{i = 1}^{d}r_{i-1}n_ir_i}{\sum_{j = 1}^{d}s_{j-1}n_js_j}
	\end{equation}
	where $r_i$ and $s_i$ are the $i$-th TT-rank of $\ten{x}\in\R^{n_1\times \dots\times n_d}$ and $\tilde{\ten{x}} = \texttt{TT-round}(\ten{x}, \delta)$. The computational cost of a \texttt{TT-round} over $\ten{x}$ in terms of floating point operations, is $\mathcal{O}(dnr^3)$, where $r = \max_{i\in\{1,\dots, d\}}\{r_i\}$ and $n = \max_{i\in\{1,\dots, d\}}\{n_i\}$, as stated in~\cite{Oseledets2011}.
	
\section{Orthogonalization schemes}\label{sec2}
	\par In the following sections, we describe the classical orthogonalization kernels and we propose their extensions to the TT-format. The input of all the orthogonalization kernels in TT-format is $\set{A}$ a set of TT-vectors and an accuracy $\delta\in\R_+$ for the TT-round function.
	
	In addition, we discuss the theoretical results for the loss of orthogonality in the classical matrix computation.
	 

	\subsection{Classical and Modified Gram-Schmidt}\label{sec2:s1}
	The Gram-Schmidt process~\cite{Gram1883, Schmidt1907} is a tool used in theoretical linear algebra to generate an orthonormal basis from a given set of vectors. Let $\set{A} =\{\vec{a}_1, \dots, \vec{a}_m\}$ be a set of $m$ linearly independent vectors of $\R^{n}$, then the key idea of the Gram-Schmidt process is to incrementally construct an orthonormal basis of the space spanned by the elements of $\set{A}$. At the $i$-th step, the $i$-th element $\vec{a}_i$ is made orthogonal to the previously computed $(i-1)$ orthonormal vectors $\{\vec{q}_1, \dots, \vec{q}_{i-1}\}$, by subtracting from $\vec{a}_i$ its projection along $\vec{q}_j$. The projection is given by the inner product of $\vec{a}_i$ and $\vec{q}_j$ for $j\in\{1, \dots, i-1\}$. After normalization the new vector is $\vec{q}_i$, the $i$-th vector of the final orthonormal basis. This mechanism is easily transported into the tensor framework. Therefore, instead of presenting the theory of the Gram-Schmidt procedure in the tensor notation, we illustrate the two different realizations of this theoretical tool only in the TT-format. We carefully emphasize the differences to the classical matrix implementations. 
	
	\subsubsection{Classical schemes without re-orthogonalization}
	The Gram-Schmidt process can be directly implemented through \emph{Classical Gram-Schmidt} (CGS), with its TT-version outlined in Algorithm~\ref{alg:CGS}. TT-CGS initializes $\ten{p}_i$ to $\ten{a}_i$ for every $i\in\{1, \dots, m\}$, see line~\ref{alg:CGS:l4}. In the core loop, the algorithm subtracts from $\ten{p}_i$ the projection of $\ten{a}_i$ along the $(i-1)$ previously computed tensors $\ten{q}_j$ of the new orthogonal basis, as described in lines~\ref{alg:CGS:l6} and~\ref{alg:CGS:l7}. Finally, $\ten{p}_i$ is normalized and added to the new orthonormal basis $\set{Q} = \{\ten{q}_1, \dots, \ten{q}_i\}$.
	The $i$-th column of $\mat{R}$ is defined using the projections of $\ten{a}_i$ along $\ten{q}_j$ for $j\in\{1,\dots, i-1\}$. By construction, $\mat{R}$ is consequently upper triangular. The norm of $\ten{p}_i$ computed in line~\ref{alg:CGS:l10} is the $i$-th diagonal entry of $\mat{R}$. These steps are present in both the tensor and matrix versions of the Classical Gram-Schmidt algorithm. However when dealing with compressed format tensors, it is important to ensure that the algorithm steps do not significantly reduce the compression quality. Therefore, it is crucial that the TT-ranks stay small. For example, after $(k-1)$ repetitions of line~\ref{alg:CGS:l7}, which involves $(k-1)$ subtractions, the TT-rank of $\ten{p}$ will be bounded by $kr$, if $r$ is the maximum TT-rank of $\ten{p}_k$ and $\ten{q}_j$ for every $j\in\{1, \dots, k-1\}$. To limit the growth of TT-rank, we compress $\ten{p}_i$ in line~\ref{alg:CGS:l9} using the \texttt{TT-round} algorithm with accuracy $\delta$. This is the most computationally expensive operation in orthogonalization algorithms. As a result, the complexity of TT-CGS depends on the number of \texttt{TT-round} calls and its complexity. This last is known to be $\bigO(dnr^3)$ where $d$ is the order of the TT-vector rounded, $n$ and $r$ are the maximum of the mode size and of the TT-rank respectively. However, in TT-CGS and in the other studied orthogonalization methods, the TT-rank is not always known. Linear combinations of TT-vectors obtained from the \texttt{TT-round} algorithm with accuracy $\delta$ are rounded, resulting in a TT-rank that is not known a priori. Therefore, we estimate the complexity of the orthogonalization algorithms here and after in terms of the number of rounding operations. The complexity of TT-CGS is equal to $m$ \texttt{TT-round} operations. 
	
	\ignore{As stated at the beginning of Section~\ref{sec2}, , we take into account only the number of times we call the TT-round algorithm, which in this case is equal to $m$.
	we assume that each TT-vector $\ten{a}_i$ and $\ten{q}_i$ has as maximum TT-rank $r$ and $r_\delta$ respectively, while the mode sizes are all bounded by $n$. The call to the \texttt{TT-round} algorithm on $\ten{p}_i$ represents the most expensive step, which is estimates as $\bigO(dnr_i^3)$ where $r_i$ is equal to the maximum TT-rank of $\ten{p}_i$. At the $i$-th iteration the maximum TT-rank of $\ten{p}_i$ is bounded by $ir_\delta$ and the \texttt{TT-round} applied to $\ten{p}_i$ cost is $\bigO(dni^3r^3_\delta)$. Since we repeat $m$ times the rounding operation, whose cost increases for $i\in\{1, \dots, m\}$, the total complexity of the TT-CGS algorithm is estimated as $\bigO(dnm^4r^3_\delta)$.}
	
	\begin{minipage}{0.46\textwidth}
		\begin{algorithm}[H]
			\centering
			\caption{$\set{Q}\,, \mat{R}$ = TT-CGS($\set{A}$, $\delta$) }\label{alg:CGS}
			\begin{algorithmic}[1]
				\Input{\small$\set{A} = \{\ten{a}_1, \dots, \ten{a}_m\}$ a set of TT-vectors, $\delta\in\R_+$ a relative rounding accuracy}
				\Output{\small$\set{Q} = \{\ten{q}_1, \dots, \ten{q}_m\}$ the set of orthogonal TT-vectors, $\mat{R}$ the upper triangular matrix}
				\For{$i = 1, \ldots, m$}
				\State $\ten{p} = \ten{a}_i$ \label{alg:CGS:l4}	
				\For{$j = 1, \ldots, i-1$}
				\LineComment\small{compute the projection of $\ten{a}_i$ along $\ten{q}_j$}
				\State $\mat{R}(i,j) = \langle \ten{a}_i,\,\ten{q}_j\rangle$\label{alg:CGS:l6} 
				\LineComment\small{remove the projection of $\ten{a}_i$ along $\ten{q}_j$}
				\State $\ten{p} = \ten{p}- \mat{R}(i,j)\ten{q}_j$\label{alg:CGS:l7} 
				\EndFor
				\State $\ten{p} = \texttt{TT-round}(\ten{p}, \delta)$\label{alg:CGS:l9}
				\State $\mat{R}(i,i) = ||\ten{p}||$\label{alg:CGS:l10}
				\State $\ten{q}_i = 1/R(i,i) \, \ten{p}$ \Comment normalize $\ten{p}$
				\EndFor
			\end{algorithmic}
		\end{algorithm}
	\end{minipage}
	\hfill
	\begin{minipage}{0.46\textwidth}
		\begin{algorithm}[H]
			\centering
			\caption{$\set{Q}\,, \mat{R}$ = TT-MGS($\set{A}$, $\delta$)}\label{alg:MGS}
			\begin{algorithmic}[1]
				\Input{\small$\set{A} = \{\ten{a}_1, \dots, \ten{a}_m\}$ a set of TT-vectors, $\delta\in\R_+$ a relative rounding accuracy}
				\Output{\small$\set{Q} = \{\ten{q}_1, \dots, \ten{q}_m\}$ the set of orthogonal TT-vectors, $\mat{R}$ the upper triangular matrix}
				
				\For{$i = 1, \ldots, m$}
				\State $\ten{p} = \ten{a}_i$
				\For{$j = 1, \ldots, i-1$}
				\LineComment\small{compute the projection of $\ten{p}$ along $\ten{q}_j$}
				\State $\mat{R}(i,j) = \langle \ten{p},\,\ten{q}_j\rangle$\label{alg:MGS:l6} 
				\LineComment\small{remove the projection of $\ten{p}$ along $\ten{q}_j$}
				\State $\ten{p} = \ten{p}- \mat{R}(i,j)\ten{q}_j$
				\EndFor
				\State $\ten{p} = \texttt{TT-round}(\ten{p}, \delta)$\label{alg:MGS:l10}				
				\State $\mat{R}(i,i) = ||\ten{p}||$
				\State $\ten{q}_i = 1/R(i,i) \, \ten{p}$ \Comment normalize $\ten{p}$
				\EndFor
				
			\end{algorithmic}
		\end{algorithm}
	\end{minipage}
	\\
	\par In the classical matrix framework, Classical Gram-Schmidt is known to suffer from a loss of orthogonality in the computed basis, as discussed later on, cf.~\cite{Bjorck1967}. The \emph{Modified Gram-Schmidt} (MGS) algorithm introduces a small algorithmic change to Classical Gram-Schmidt, which guarantees better numerical orthogonality. In TT-MGS, we remove the projection of $\ten{p}_i$, rather than of $\ten{a}_i$, along $\ten{q}_j$ for every $j\in\{1, \dots, i-1\}$. This can be seen by comparing line~\ref{alg:CGS:l6} of Algorithm~\ref{alg:CGS} and~\ref{alg:MGS} respectively. This modification reduces error propagation, improving the algorithm's general stability in both the classical matrix and tensor cases, as we discussed in Section~\ref{sec2:s4}. The remaining steps of the two algorithms are identical.
	
	\ignore{Therefore after removing from $\ten{p}_i$ its projection along $\ten{q}_j$ for $j\in\{1, \dots, i-1\}$, it is rounded with accuracy $\delta$ to prevent memory deficiency, normalized and added to the set $\set{Q} = \{\ten{q}_1, \dots, \ten{q}_i\}$.} Under the same assumptions stated previously to estimate the complexity, we conclude that TT-MGS computational complexity in TT-format is equal to TT-CGS one, given by $m$ \texttt{TT-round} calls.
	\subsubsection{Classical schemes with re-orthogonalization}
	CGS and MGS are known to have stability issues, as described in detail in Section~\ref{sec2:s4}. The closer the input vectors are to linear dependence, the more the algorithms propagate the rounding errors, spoiling the final orthogonality of the new basis. As reported in~\cite{Giraud2005}, several articles, such as~\cite{Abdelmalek1971, Daniel1976, Hoffmann1989}, have addressed this issue by introducing re-orthogonalization steps. This involves repeatedly orthogonalizing the basis using the same approach. In~\cite{Giraud2005}, the authors showed theoretically that one re-orthogonalization step is sufficient to significantly improve the orthogonality of the new basis generated by CGS and MGS. We briefly introduce the concepts of CGS and MGS with re-orthogonalization, referred to as TT-CGS2 and TT-MGS2, respectively, in the tensor case. We emphasize the steps that are unique to the TT-version.
	\\
	\begin{minipage}{0.46\textwidth}
		\begin{algorithm}[H]
			\centering
			\caption{$\set{Q}\,, \mat{R}$ = TT-CGS2($\set{A}$, $\delta$) }\label{alg:CGS2}
			\begin{algorithmic}[1]
				\Input{\small$\set{A} = \{\ten{a}_1, \dots, \ten{a}_m\}$ a set of TT-vectors, $\delta\in\R_+$ a relative rounding accuracy}
				\Output{\small$\set{Q} = \{\ten{q}_1, \dots, \ten{q}_m\}$ the set of orthogonal TT-vectors, $\mat{R}$ the upper triangular matrix}
				
				\For{$i = 1, \ldots, m$}
				\State $\ten{p}_0 = \ten{a}_i$
				\LineComment\footnotesize{repeat twice the orthogonalization loop}
				\For{$k = 1,2$}\label{alg:CGS2:l4}
				\State $\ten{p}_k = \ten{p}_{k-1}$ 
				\For{$j = 1, \ldots, i-1$}\label{alg:CGS2:l6}
				\LineComment\footnotesize{compute the projection of $\ten{p}_{k-1}$ along $\ten{q}_j$}
				\State $\mat{R}_k(i,j) = \langle \ten{p}_{k-1},\,\ten{q}_j\rangle$\label{alg:CGS2:l7}
				\LineComment\footnotesize{subtract the projection of $\ten{p}_{k-1}$ along $\ten{q}_j$}
				\State $\ten{p}_k = \ten{p}_k- \mat{R}_k(i,j)\ten{q}_j$ 
				\EndFor
				\State $\ten{p}_k = \texttt{TT-round}(\ten{p}_k, \delta)$\label{alg:CGS2:l10}
				\EndFor
				\State $\mat{R}_2(i,i) = ||\ten{p}_2||$ \label{alg:CGS2:l12}
				\State $\ten{q}_i = 1/R_2(i,i)\ten{p}_2$\label{alg:CGS2:l13}\Comment normalize $\ten{p}_2$
				\EndFor
				\LineComment\footnotesize{compute the R factor from the repeated orthogonalization loop}
				\State $\mat{R} = \mat{R}_1 + \mat{R}_2$
			\end{algorithmic}
		\end{algorithm}
	\end{minipage}
	\hfill
	\begin{minipage}{0.46\textwidth}
		\begin{algorithm}[H]
			\centering
			\caption{$\set{Q}\,, \mat{R}$ = TT-MGS2($\set{A}$, $\delta$) }\label{alg:MGS2}
			\begin{algorithmic}[1]
			\Input{\small$\set{A} = \{\ten{a}_1, \dots, \ten{a}_m\}$ a set of TT-vectors, $\delta\in\R_+$ a relative rounding accuracy}
			\Output{\small$\set{Q} = \{\ten{q}_1, \dots, \ten{q}_m\}$ the set of orthogonal TT-vectors, $\mat{R}$ the upper triangular matrix}
			
				\For{$i = 1, \ldots, m$}
				\State $\ten{p}_0 = \ten{a}_i$
				\LineComment\footnotesize{repeat twice the orthogonalization loop}
				\For{$k = 1,2$}
				\State $\ten{p}_k = \ten{p}_{k-1}$ 
				\For{$j = 1, \ldots, i-1$}
				\LineComment\footnotesize{compute the projection of $\ten{p}_{k}$ along $\ten{q}_j$}
				\State $\mat{R}_k(i,j) = \langle \ten{p}_{k},\,\ten{q}_j\rangle$
				\LineComment\footnotesize{subtract the projection of $\ten{p}_{k}$ along $\ten{q}_j$}
				\State $\ten{p}_k = \ten{p}_k- \mat{R}_k(i,j)\ten{q}_j$
				\EndFor
				\State $\ten{p}_k = \texttt{TT-round}(\ten{p}_k, \delta)$
				\EndFor
				\State $\mat{R}_2(i,i) = ||\ten{p}_2||$ 
				\State $\ten{q}_i = 1/R_2(i,i)\ten{p}_2$\Comment normalize $\ten{p}_2$
				\EndFor	
				\LineComment\footnotesize{compute the R factor from the repeated orthogonalization loop}
				\State $\mat{R} = \mat{R}_1 + \mat{R}_2$
			\end{algorithmic}
		\end{algorithm}
	\end{minipage}
	\vspace{10pt}
	\par In CGS2, described in Algorithm~\ref{alg:CGS2}, the input TT-vector $\ten{a}_i$ is orthogonalized with respect to the previously computed orthogonal TT-vectors $\{\ten{q}_1, \dots, \ten{q}_{i-1}\}$, by subtracting from $\ten{a}_i$ its projection along $\ten{q}_j$. The projection of $\ten{a}_i$ along $\ten{q}_j$ defines the $(j,i)$ element of the first matrix $R_1$. These first $(i-1)$ iterations, given in line~\ref{alg:CGS2:l6} of Algorithm~\ref{alg:CGS2}, define the TT-vector $\ten{p}_1$. Then, in line~\ref{alg:CGS2:l10}, $\ten{p}_1$ is rounded. Up to this point, TT-CGS2 functions identically to TT-CGS. However, in TT-CGS2, the TT-vector $\ten{p}_1$ is orthogonalized again against $\{\ten{q}_1, \dots, \ten{q}_{i-1}\}$, after rounding. This results in $\ten{p}_2$. The projections along $\ten{q}_j$ of $\ten{p}_1$ determine the $(j,i)$ component of the second matrix $R_2$. Once $\ten{p}_2$ is fully defined, it is rounded and normalized, defining the $i$-th orthogonal TT-vector $\ten{q}_i$, as stated in lines~\ref{alg:CGS2:l12} and~\ref{alg:CGS2:l13} of Algorithm~\ref{alg:CGS2}. The $\ten{p}_2$ norm at the $i$-th iteration determines the $(i,i)$-th diagonal component of $R_2$. The R factor from the QR decomposition computed by CGS2 is obtained by adding $R_1$ and $R_2$. The distinction between MGS2 and CGS2 is evident in line~\ref{alg:CGS2:l7} of Algorithm~\ref{alg:MGS2} and~\ref{alg:CGS2}, respectively. In the first orthogonalization loop, that is, when $k=1$ in line~\ref{alg:CGS2:l4} of both methods, the classical Gram-Schmidt version projects $\ten{a}_i$ is along $\ten{q}_j$ defining $\ten{p}_1$, while the modified Gram-Schmidt version projects $\ten{p}_1$ along $\ten{q}_j$ to update it. In the second orthogonalization loop, i.e., when $k=2$ in line~\ref{alg:CGS2:l4} of both algorithms, TT-CGS2 removes the projection of $\ten{p}_1$ along $\ten{q}_j$ to define $\ten{p}_2$. In the TT-MGS2 version, $\ten{p}_2$ is the TT-vector projected along $\ten{q}_j$ and updated. The remaining steps, including the \texttt{TT-round} and the construction of $R_1$, $R_2$ and their sum $R$, are identical between TT-CGS2 and TT-MGS2. It is important to note that the rounding steps are only performed in the TT-version of MGS2 and CGS2.
	\par The computational complexity of TT-CGS2 and TT-MGS2 is estimated as $2m$ \texttt{TT-round} operations, based on the previously used hypothesis. During the $m$ iterations, the two temporary TT-vectors $\ten{p}_1$ and $\ten{p}_2$ are rounded.

	\subsection{Gram approach}\label{sec2:s2}
	In their article~\cite{Stathopoulos2002}, the authors propose an algorithm for generating an orthogonal basis from a set of $m$ linearly independent vectors of $\R^{n}$ with $m \ll n$. We will refer to this algorithm as Gram's algorithm. This scheme is based on the Gram matrix, which under the hypothesis that $m$ is significantly small. The key idea is to decompose the small Gram matrix by its Cholesky factorization and use it to generate the orthogonal basis. We briefly describe the main ideas of this orthogonalization scheme in the classical matrix framework and we describe in detail the implementation of the Gram algorithm in the TT-format. In fact, the tensor realization of this procedure is extremely close to the matrix one, so a description of the tensor case and its differences from the classical matrix one is sufficient to ensure a good understanding.
	\par Given a set of vectors $\set{A} = \{\vec{a}_1, \dots, \vec{a}_m\}$ with $\vec{a}_i\in\R^{n}$, the Gram matrix $\mat{G}\in\R^{m\times m}$ is defined by the inner product $\mat{G}(i,j) = \langle \vec{a}_i,\, \vec{a}_j\rangle$ for every $i,j\in\{1,\dots, m\}$. Equivalently, let $\vec{a}_i$ be the $i$-th column of the matrix $\mat{A}\in\R^{n\times m}$, then in the matrix computation the Gram matrix is written as 
	\begin{equation}
	\label{eq:GRAM:A}
	\mat{G} = \mat{A}^\top \mat{A}.
	\end{equation}
	If the elements of $\set{A}$ are linearly independent, then $\mat{G}$ is symmetric positive definite. As a consequence, its Cholesky factorization exists and is written as $\mat{G} = \mat{L}\mat{L}^\top$, where $\mat{L}\in\R^{m\times m}$ is a lower triangular matrix. If we now denote the transpose of $\mat{L}$ by $\mat{R}$, then the Gram matrix gets 
	\begin{equation}
	\label{eq:GRAM:R}
	\mat{G} = \mat{R}^\top \mat{R}.
	\end{equation}
	Comparing Equations~\eqref{eq:GRAM:A} and~\eqref{eq:GRAM:R}, we conclude that $\mat{R}$ is the R-factor from the QR decomposition of $A$, i.e., it expresses the same information of $\mat{A}$ in a different basis. The matrix $Q$ from the QR decomposition of $A$ is written as $\mat{Q} = \mat{A}\mat{R}^{-1}$ where $\mat{R} = \mat{L}^\top$. The columns of $\mat{Q}$ form an orthogonal basis $\set{Q} = \{\vec{q}_1, \dots, \vec{q}_m\}$, whose $j$-th element is strictly speaking a linear combination of the first $j$ elements of $\set{A}$, i.e.,
	\begin{equation*}
	\vec{q}_j = \sum_{k=1}^{j}\mat{R}^{-1}(k,j)\vec{a}_k.
	\end{equation*}
	
	\begin{remark}
		Note that, by construction, the condition number of $\mat{G}$ is the square of that of $\mat{A}$. Consequently, if the condition number of $\mat{A}$ associated with the set of input vectors $\set{A}$ is greater than the inverse of the square root of the working precision of the arithmetic considered , e.g., $u_{64}\approx 10^{-16}$ for 64-bit computation, the associated Gram matrix $\mat{G}$ is numerically singular and its Cholesky decomposition is no longer defined. This is the main practical drawback of this method. 
	\end{remark}
	
	This procedure generates an orthonormal basis starting from a set of linear independent vectors, which is naturally extended to TT-vectors. As described in Algorithm~\ref{alg:GRAM}, given a set of TT-vectors $\set{A} = \{\ten{a}_1, \dots, \ten{a}_m\}$ with $\ten{a}_i\in\R^{n_1\times \dots \times n_d}$, we construct the Gram matrix $G\in\R^{m\times m}$ by the tensor inner product and we compute its Cholesky factorization to obtain the lower triangular matrix $L\in\R^{m\times m}$. As in the matrix case, $\mat{R}\in\R^{m\times m}$ the transpose of $\mat{L}$ expresses the same information as the TT-vectors of $\set{A}$, but with respect to a different basis. Following the matrix approach, we retrieve this basis, i.e., the orthonormal set $\set{Q}\,$ whose element $\ten{q}_i\in\R^{n_1\times \dots \times n_d}$ is defined as
	\begin{equation*}
	\ten{q}_i = \sum_{k=1}^{i}R^{-1}(k,i)\ten{a}_k.
	\end{equation*} 
	\begin{remark}
		In the matrix framework, the orthogonal vector $\vec{q}_j$ is obtained from the elements of $\set{A}$ by back-substitution using the matrix $\mat{R}$. However, this approach does not easily translated to the tensor framework, where the inverse of $R$ must be explicitly computed.
	\end{remark}
	As for the other orthogonalization techniques, we avoid memory problems by monitoring the TT-ranks and eventually rounding. Indeed, assuming that all the TT-vectors of $\set{A}$ have TT-ranks bounded by $r$, then the $i$-th TT-vector constructed in line~\ref{alg:GRAM:l6} has a maximum TT-rank bounded by $ir$. Since this value grows linearly with $m$, in line~\ref{alg:GRAM:l7} we introduce a rounding step with prescribed accuracy $\delta$. 
	\begin{algorithm}[htb]
		\centering
		\caption{$\set{Q}\,, \mat{R}$ = TT-Gram($\set{A}$, $\delta$)}\label{alg:GRAM}
		\begin{algorithmic}[1]
			\Input{$\set{A} = \{\ten{a}_1, \dots, \ten{a}_m\}$ a set of TT-vectors, $\delta\in\R_+$ a relative rounding accuracy}
			\Output{$\set{Q} = \{\ten{q}_1, \dots, \ten{q}_m\}$ the set of orthogonal TT-vectors, $\mat{R}$ the upper triangular matrix}
			
			\For{$i = 1, \dots, m$}
			\For{$j = 1, \dots, i$}
			\LineComment construct the Gram matrix through the inner product of the input TT-vectors
			\State$\mat{G}(i,j) = \mat{G}(j,i) = \langle \ten{a}_i,\,\ten{a}_j\rangle$
			\EndFor
			\EndFor
			\State $\mat{L} = \texttt{cholesky}(\mat{G})$ \Comment compute the Cholesky factorization
			\State $\mat{R} = \mat{L}^\top$ and $\mat{R}^{-1} = \texttt{invert}(R)$\Comment define the R factor of the QR-factorization
			\For{$i = 1, \dots, m$}
			\State $\ten{p} = \mat{R}^{-1}(i,1)\ten{a}_1$
			\For{$j = 2, \dots, i$}
			\LineComment construct the $i$-th new basis TT-vector as a linear combination of the $(i-1)$ input TT-vector
			\State $\ten{p} = \ten{p} +\mat{R}^{-1}(i,j)\ten{a}_j$	\label{alg:GRAM:l6} 
			\EndFor
			\State $\ten{q}_i = \texttt{TT-round}(\ten{p}, \delta)$\label{alg:GRAM:l7}	\Comment round the TT-vector before adding it to the basis
			\EndFor
		\end{algorithmic}
	\end{algorithm}
	As in Sections~\ref{sec2:s1} and~\ref{sec2:s2}, note that the complexity of the TT-Gram algorithm is given by $m$ \texttt{TT-round} operations. However, in this particular case, we can even estimate the cost of each single rounding step, and thus of the entire algorithm. Indeed, the maximum TT-rank of the rounded TT-vector $\ten{q}_i$ is bounded by $i\,r$, under the assumption that the maximum TT-rank and the maximum mode size of $\ten{a}_i$ are bounded by $r\in\N$ and $n\in\N$ respectively. Consequently, the computational cost, i.e., the number of floating point operations, of each rounding operation, the most expensive step in the entire algorithm, is known and is equal to $\bigO(dni^3r^3)$; summing over $i\in\{1,\dots, m\}$, we conclude that the cost of the TT-Gram algorithm is $\bigO(dnm^4r)$ floating point operations.
	
	\ignore{To estimate the complexity of the TT-Gram algorithm, we the. As in the previously analysed cases, the most expensive operation is the \texttt{TT-round} in line~\ref{alg:GRAM:l7} whose cost depends on the TT-rank of $\ten{p}$ for line~\ref{alg:GRAM:l6}. At iteration $i$, the TT-rank of $\ten{p}$ is bounded by $ir$ and as consequence the rounding cost is . The final cost of the TT-Gram algorithm is . Differently from TT-CGS and TT-MGS, in this case the algorithm complexity is independent from the rounding accuracy chosen.}
	
	\subsection{Householder reflections}\label{sec2:s3}
	In the classical matrix framework, Householder transformations are commonly used to generate an orthogonal basis due to their stability properties, as explained in the following sections. We will briefly present the theoretical construction of a Householder transformation and how it can be used to generate an orthogonal basis. The following section provides a detailed description of how the Householder transformation is extended to the tensor context, with specific attention given to the implementation of the Householder orthogonalization scheme in TT-format. 
	
	\par The Householder reflector is used to move a vector $\vec{x}\in\R^{n}$ along a chosen direction, which is typically an element of the canonical basis or a linear combination of them. The construction of the Householder reflector in the general case is illustrated. Let $\vec{x}\in\R^{n}$ be the vector we want to reflect along the normalized vector $\vec{y}\in\R^{n}$, the \emph{Householder reflection or transformation} is a linear operator $\op{H}:\R^n\rightarrow \R^{n}$ such that
	\begin{equation*}
	\op{H}(\vec{x}) = \norm{\vec{x}}\vec{y}\qquad\text{with}\qquad\norm{y}=1.
	\end{equation*}
	The Householder reflector is represented with respect to the canonical basis of $\R^{n}$ by the matrix $\mat{H}\in\R^{n\times n}$ such that $\mat{H} = \I_n - 2\vec{u}\otimes \vec{u}$ where $\vec{u}\in\R^{n}$ is the \emph{Householder vector} defined as
	\begin{equation}
	\label{eq:HH:vec}
	\vec{u} = \frac{\vec{x} - \vec{z}}{\norm{\vec{x} - \vec{z}}}\qquad\text{with}\qquad \vec{z} = \norm{\vec{x}}\vec{y}.
	\end{equation}
	The Householder reflection matrix $\mat{H}$ is unitary and defined entirely by the Householder vector $\vec{u}$. Additionally, the action of a Householder reflector is computed by one inner product with the Householder vector $\vec{u}$ and one algebraic vector summation. For a given vector $\vec{w}\in\R^{n}$ and a Householder reflector $\mat{H} = \I_n - 2\vec{u}\otimes \vec{u}$, the image of $\vec{w}$ through $\mat{H}$ is
	\begin{equation}
	\label{eq:HH:action}
	\mat{H}\vec{w} = \vec{w} - 2\langle\vec{w},\,\vec{u}\rangle\vec{u}.
	\end{equation} 
	If $\vec{u}$ is defined as in Equation~\eqref{eq:HH:vec}, then it can be verified that $\mat{H}\vec{x} = \norm{x}\vec{y}$. It is important to note that
	\begin{equation*}
	\begin{split}
	\norm{x - z}^2 &= \langle \vec{x}-\norm{\vec{x}}\vec{y},\,\vec{x}-\norm{\vec{x}}\vec{y}\rangle \\
	&= \norm{x}^2 -2\norm{x}\langle \vec{x}, \, \vec{y}\rangle + \norm{x}^2\norm{y}^2 \\
	&= 2\bigl(\norm{x}^2 -\norm{x}\langle \vec{x}, \, \vec{y}\rangle\bigr)
	\end{split}
	\end{equation*}
	since $\norm{y} = 1$ by hypothesis. Using this result and Equation~\eqref{eq:HH:action}, we can obtain 
	\begin{equation*}
	\begin{split}
	\mat{H}\vec{x} &= \vec{x} - 2\langle\vec{x},\,\vec{u}\rangle\vec{u} \\
	&= \vec{x} - \frac{2}{2\bigl(\norm{x}^2 -\norm{x}\langle \vec{x}, \, \vec{y}\rangle \bigr)}\langle\vec{x},\,\vec{x}-\norm{\vec{x}}\vec{y}\rangle(\vec{x}-\norm{\vec{x}}\vec{y})\\
	&= \vec{x} - \frac{1}{\bigl(\norm{x}^2 -\norm{x}\langle \vec{x}, \, \vec{y}\rangle \bigr)}\bigl(\norm{x}^2 - \norm{x}\langle \vec{x}, \, \vec{y}\rangle \bigr)(\vec{x}-\norm{\vec{x}}\vec{y})\\
	&= \vec{x} - (\vec{x}-\norm{\vec{x}}\vec{y})\\
	&=\norm{\vec{x}}\vec{y}
	\end{split}
	\end{equation*}
	\par Householder transformations are commonly used to compute the QR factorization of a matrix, but they can also be applied when a set of vectors needs to be converted into an orthogonal basis. We will briefly examine the two possibilities. When given a matrix $\mat{A}\in\R^{n\times m}$, we construct $m$ Householder reflections. The $k$-th reflection moves the $k$-th column of $\mat{A}$ along a linear combination of the first $k$ canonical basis vectors. In other words, the $k$-th Householder transformation sets the last $(n-k)$ entries of the $k$-th column of $\mat{A}$ to zero. As a result, after $m$ Householder reflections, the matrix $\mat{A}$ becomes upper triangular. We will now provide a more detailed illustration of how the algorithm iteratively proceeds. To begin with, let $\vec{a}_1\in\R^{n}$ represent the first column of $\mat{A}$. The first step is to reflect it along the first canonical basis vector $\vec{e}_1$ by constructing the Householder reflector $H_1$ such that $H_1\vec{a}_1 = \norm{\vec{a}_1}\vec{e}_1$. The first Householder vector $\vec{u}_1\in\R^{n}$ is then
	\begin{equation*}
	\vec{u}_1 = \vec{a}_1 \pm \norm{\vec{a}_1}\vec{e}_1
	\end{equation*} 
	and normalized.
	For stability reasons (cf.~\cite{trefethen97}), the sign of the norm of $\vec{a}_1$ is determined by the sign of the first component of $\vec{a}_1$, which is positive if $\vec{a}_1(1) > 0$ and negative otherwise. The first Householder reflector $H_1$ is applied to all the columns of $\mat{A}$, resulting in $\tilde{a}_j = \mat{H}_1\vec{a}_j$ for $j\in\{1, \dots, m\}$. From now on, $\tilde{\vec{a}}_j$ denotes the $j$-th column of $\mat{A}$ updated by all the previously defined $(j-1)$ Householder transformations for $j\in\{2, \dots, m\}$. It is important to note that the first Householder transformation moves the first column of $\mat{A}$ along a multiple of the first canonical basis vector $e_1$, i.e., setting the last $(n-1)$ entries of the first column of $\mat{A}$ to zero. Next, we reflect the second column of $\mat{A}$ along a linear combination of the first two canonical basis vectors $\vec{e}_1$ and $\vec{e}_2$. We define $\vec{u}_2\in\R^{n}$ as the Householder vector that defines the second Householder reflector $H_2$. This second Householder reflector updates the $j$-th column of $\mat{A}$ for $j\in\{2, \dots, m\}$ a second time. At this point, only the first two entries of $\tilde{\vec{a}}_2$ are different from zero. The $k$-th Householder reflection $H_k$ moves $\tilde{\vec{a}}_{k}\in\R^{n}$ the $k$-th column of $A$, updated by the first $(k-1)$ Householder reflections along a linear combination of the first $k$ elements of the canonical basis of $\R^{n}$, i.e., $\mat{H}_k\tilde{\vec{a}}_{k} = \sum_{\ell=1}^{k}\alpha_\ell \vec{e}_\ell$ with $ \sqrt{\sum_{\ell=1}^{k} \alpha^2_\ell} = ||\tilde{\vec{a}}_{k}||$. Prior to normalization, the $k$-th Householder vector, $\vec{u}_{k}\in\R^{n}$, is defined as
	\begin{equation}
	\label{eq:Hvec}
	\vec{u}_{k} = \tilde{\vec{a}}_{k} - \sum_{\ell=1}^{k}\beta_\ell\vec{e}_\ell,
	\end{equation}
	where 
	\begin{equation*}
	\beta_\ell = \tilde{\vec{a}}_{k}(\ell) \qquad\text{and}\qquad \beta_{k} = \pm\sqrt{||\tilde{\vec{a}}_{k}||^2 - \sum_{\ell=1}^{k-1} \beta_\ell^2},
	\end{equation*}
	for every $\ell\in\{1, \dots, k-1\}$.
	To ensure stability (cf.~\cite{trefethen97}), $\beta_{k}$ is positive if $\tilde{\vec{a}}_k(k)$ is positive, and negative otherwise. The vector $\vec{u}_k$ is then normalized.
	\begin{remark}
		\label{rk:HH}
		By construction, the first $(k-1)$ entries of $\vec{u}_k$ are zeros, the last $(n-k)$ entries are equal to the corresponding ones of $\tilde{\vec{a}}_k$. The $k$-th component of $\vec{u}_k$ is obtained by subtracting the quantity $\beta_k$ from the $k$-th component of $\vec{a}_k$, that is $\vec{u}_k(k) = \tilde{\vec{a}}_k(k) - \beta_{k}$. In the context of matrices, the Householder QR factorization has a simplified construction due to the property that only the last $(n-k+1)$ entries of $\tilde{\vec{a}}_k$ have a determinant role. The $k$-th Householder vector, $\hat{\vec{u}}_k\in\R^{n-k+1}$, is defined from the norm of $\tilde{a}_k$, with the first $(k-1)$ entries being zeros. In other words,
		\begin{equation*}
		\hat{\vec{u}}_k = \gamma_k\vec{e}_1\qquad\text{where}\qquad\gamma_k =\sqrt{ \sum_{{j}=k}^{n}\bigl(\tilde{\vec{a}}_k(j)\bigr)^2}
		\end{equation*}
		with $e_1\in\R^{n-k+1}$ for every $k\in\{1, \dots, m\}$. The $k$-th Householder transformation, $H_k\in\R^{(n-k+1)\times (n-k+1)}$, is defined based on $\hat{\vec{u}}_k$ and is only applied on the last $(n-k+1)$ components of $\tilde{a}_j$, where $j\in\{k, \dots, m\}$ and $k\in\{1, \dots, m\}$. It is important to note that this reduced approach cannot be replicated in the tensor framework, where objects are expressed in compressed format. Therefore, we present it in the most general way, which is not the same as the approach used for in matrix computation.
	\end{remark}
	
	After applying the $k$-th Householder transformation to $\tilde{\vec{a}}_j$ for $j\in\{k, \dots, m\}$, the vector $\tilde{\vec{a}}_k$ will have its last $(n-k)$ component equal to zero. The application of $m$ Householder reflections to $\mat{A}$ results in the upper triangular matrix $R$, which is the R factor of the QR decomposition. To calculate the Q factor, we multiply the $m$ Householder reflection matrices. For most applications, it is sufficient to form the Householder vectors and know the Householder transformations implicitly, as given in Equation~\eqref{eq:HH:action}. However, if we want to produce an orthogonal basis from a generic set of $m$ vectors $\set{A} = \{\vec{a}_1, \dots, \vec{a}_m\}$ with $\vec{a}_k\in\R^{n}$, we must take an additional step. When computing the QR factorization, the $k$-th Householder transformation, $\mat{H}_k$, is defined by the $k$-th Householder vector, $\vec{u}_{k}$, as given in Equation~\eqref{eq:Hvec} for every $k\in\{1, \dots, m\}$. To generate the set of orthonormal vectors $\set{Q} = \{\vec{q}_1, \dots, \vec{q}_m\}$, we apply the first $k$ Householder transformations, $\mat{H}_k$, to the $k$-th canonical basis vector, $\vec{e}_k$, in reverse order. That is
	\begin{equation*}
	\vec{q}_k = \mat{H}_1\cdots\mat{H}_k\vec{e}_k.
	\end{equation*} 
	\par We extend this approach to the tensor case with some modifications, as in the previous sections. Let $\set{A}$ be a set of $m$ TT-vectors $\ten{a}_i\in\R^{n_1\times \dots \times n_d}$ for every $i\in\{1, \dots, m\}$. To construct the Householder transformations, we first define a \emph{canonical} basis for a tensor subspace of dimension $m$ of $\R^{n_1\times \dots \times n_d}$. In order to do so, we fix $\set{N}\;_i = \{1,\dots, n_i\}$ for every $i\in\{1,\dots, d\}$ and $n = \prod_{j=1}^d n_j$. We then define the function $\psi: \set{N}\;_1\times\dots\times\set{N}\;_d\rightarrow\{1,\dots, n\}$ such that
	\begin{equation*}
	\psi(i_1,\dots, i_d) = i_1 + \sum_{\alpha = 2}^{d}(i_\alpha - 1)m_\alpha\qquad\text{with}\qquad m_\alpha = \prod_{\beta = 1}^{\alpha-1}n_\beta.
	\end{equation*} 
	Since $\psi$ is invertible, we denote its inverse by $\phi:\{1,\dots, n\}\rightarrow \set{N}\;_1\times\dots\times\set{N}\;_d$ such that $\phi(i) = (i_1, \dots, i_d)$. As a consequence, $\psi(\phi(i)) = i$ and $\phi(\psi(i_1,\dots, i_d)) = (i_1,\dots, i_d)$ for $i\in\{1,\dots, n\}$ and $i_k\in\{1,\dots, n_k\}$ with $k\in\{1,\dots, d\}$.
	The basis for the subspace of dimension $m$ of $\R^{n_1\times \dots \times n_d}$ is fixed as $\set{E} = \{\ten{e}_1, \dots, \ten{e}_m\}$ with 
	\begin{equation}
	\ten{e}_i = e_{i_1}\otimes\cdots\otimes e_{i_d} \qquad\text{with} \qquad ({i_1,\dots, i_d}) = \phi(i),
	\end{equation}
	where $e_{i_k}$ is the $i_k$-th canonical basis vector of $\R^{n_k}$ for $k\in\{1,\dots, d\}$. The index $i$ is used to denote the $i$-th element of the canonical basis, that is $i = \psi(i_1, \dots, i_d)$.
	\par As stated in Remark~\ref{rk:HH}, the vector $\vec{u}_k$ has zeros for its first $(k-1)$ components, the corresponding components of $\tilde{\vec{a}}_k$ for its last $(n-k)$ entries, and the difference between the $k$-th component of $\tilde{\vec{a}}_k$ and the quantity $\beta_k$ (defined in Equation~\eqref{eq:Hvec}) for its $k$-th component. This structure needs to be transported in the tensor case. However, if the elements of $\set{A}$ are in TT-format, it is not possible to directly access the tensor components. We need to recover them either by multiplying the TT-cores with the correct index or by computing the inner product with the element of $\set{E}$. The $k$-th Householder TT-vector is $\ten{u}_{k}\in\R^{n_1\times \dots \times n_d}$ defined as
	\begin{equation*}
	\ten{u}_{k} = \tilde{\ten{a}}_{k} - \sum_{j=1}^{k}\mat{R}(j,k)\ten{e}_j
	\end{equation*}
	where $\tilde{\ten{a}}_{k}$ is the result of $(k-1)$ Householder reflections applied to $\ten{a}_{k}$, 	$\mat{R}(j,k) = \langle \tilde{\ten{a}}_{k},\, \ten{e}_j\rangle$ for every $j\in\{1, \dots, k\}$ and $\mat{R}(k,k) =\pm \sqrt{||\tilde{\ten{a}}{(k)}_{k}||^2 - \sum_{\ell=1}^{k-1} \mat{R}(\ell,k)^2}$ as described in lines~\ref{alg:HH-vec:l3} and~\ref{alg:HH-vec:l6} of Algorithm~\ref{alg:HH-vec}, respectively. The $k$-th component of $\vec{r}_{k}$ takes a positive sign if $\langle \tilde{\ten{a}}_{k},\,\ten{e}_{k}\rangle > 0$; otherwise, it takes negative sign. This extends the stability preserving idea given in~\cite{trefethen97} to the tensor framework. The $j$-th component of $\vec{r}_k$ corresponds to the $(j,k)$ component of the R factor. As previously mentioned, it is important to ensure that the TT-rank of $\ten{u}_k$ remains small for every $k\in\{1, \dots, m\}$. Assuming that the maximum TT-rank of $\tilde{\ten{a}}_{k}$ is bounded by $r_{\vec{a}}$, then after removing the first $(k-1)$ components of $\tilde{\ten{a}}_k$ (after line~\ref{alg:HH-vec:l3}), the TT-rank of $\ten{u}_k$ is bounded by $(r_{\vec{a}} + k-1)$. At this step, we make the first \texttt{TT-round} call on the Householder TT-vector $\ten{u}_k$, whose TT-rank decreases depending on the accuracy value $\delta$. Then we subtract the $k$-th component of $\vec{r}_k$, which results in further growth in the TT-rank of $\ten{u}_k$. As the $\ten{u}_k$ TT-vector plays a crucial role in the Householder transformation and its TT-rank has a significant impact on the entire process, we perform an additional \texttt{TT-round} step over $\ten{u}_k$ at accuracy $\delta$.
	\begin{algorithm}[htb]
		\centering
		\caption{$\ten{u}, \vec{r} = \texttt{TTH-vec}(\ten{a}, \set{F}, \delta$)}\label{alg:HH-vec}
		\begin{algorithmic}[1]
			\Input{$\ten{a}\in\R^{n_1\times \dots\times n_d}$ a TT-vector, $\set{F} = \{\ten{f}_1, \dots, \ten{f}_i\}$ a subset of the canonical tensor space basis in TT-format, $\delta\in\R_+$ a relative rounding accuracy}
			\Output{$\ten{u}$ the Householder TT-vector, $\vec{r}$ a column of the R-factor}
			\State $s = 0$
			\State $\ten{w} = \ten{a}$
			\For{$j = 1, \dots, i-1$}
			\LineComment compute the component of $\ten{a}$ along the $j$-th canonical basis TT-vector
			\State $\vec{r}(j) = \langle\ten{a},\,\ten{f}_j \rangle$ 
			\State $s = s + \bigl(\vec{r}(j)\bigr)^2$ 
			\LineComment set to zero the component of $\ten{a}$ along the $j$-th canonical basis TT-vector $\ten{f}_j$
			\State $\ten{w} = \ten{w} - \vec{r}(j)\ten{f}_j$
			\EndFor\label{alg:HH-vec:l3}
			
			\State $\ten{w} = \texttt{TT-round}(\ten{w}, \delta)$\label{alg:HH-vec:l5}
			\LineComment subtract from the norm of $\ten{a}$ the contribution of the components set to zero
			\State $\vec{r}(i) = \text{sign}(\langle\ten{a},\,\ten{f}_i
			\rangle)\sqrt{||\ten{a}||^2 - s}$\label{alg:HH-vec:l6}
			\State $\ten{w} = \ten{w} - \vec{r}(i)\ten{f}_i$
			\State $\ten{w} = \texttt{TT-round}(\ten{w}, \delta)$
			\State $\ten{u} = (1/||\ten{z}||)\ten{w}$
			
		\end{algorithmic}
	\end{algorithm}
	
	After describing the construction of a Householder TT-vector, which is summarized in Algorithm~\ref{alg:HH-vec}, the focus shifts on generation of the orthonormal TT-vector set from a generic TT-vector set $\set{A} = \{\ten{a}_1, \dots, \ten{a}_m\}$, as depicted in Algorithm~\ref{alg:HH}. To reflect the $k$-th TT-vector of $\set{A}$ along a linear combination of the first $k$ elements of $\set{E}$, we generate the $k$-th Householder TT-vector, $\ten{u}_k$, using Algorithm~\ref{alg:HH-vec}. This TT-vector defines the Householder transformation, $\ten{H}_k$, implicitly. The first $k$ components of $\vec{r}_k$ are stored in the $k$-th column of the upper triangular matrix, $\mat{R}\in\R^{m\times m}$. We apply $\ten{H}_k$ implicitly using $\ten{u}_k$ to form $\tilde{\ten{a}}_j$ for every $j\in\{k, \dots, m\}$, as expressed in line~\ref{alg:HH:l7} of Algorithm~\ref{alg:HH}. This follows the same approach as in the matrix case, where 
	\begin{equation*}
	\ten{H}_k(\tilde{\ten{a}}_j) = \tilde{\ten{a}}_j - 2\langle \tilde{\ten{a}}_j,\,\ten{u}_k\rangle\ten{u}_k.
	\end{equation*} 
	Algorithm~\ref{alg:HH_ap_ref} applies a given Householder reflection to a specific input vector. 
	\begin{algorithm}[htb]
		\centering
		\caption{$\ten{b} = \texttt{apply-H-vec}(\ten{a}, \ten{u})$}\label{alg:HH_ap_ref}
		\begin{algorithmic}[1]
			\Input{$\ten{a}\in\R^{n_1\times \dots\times n_d}$ a TT-vector to project, $u\in\R^{n_1\times \dots\times n_d}$ the Householder TT-vector}
			\Output{$\ten{b}$ the TT-vector resulting from the Householder reflector defined by $\ten{u}$ applied to $\ten{a}$}
			\State $\ten{b} = \ten{a} - 2\langle \ten{a},\, \ten{u}\rangle \ten{u}$	\Comment apply the Householder reflection defined by $\ten{u}$ to $\ten{a}$
			
		\end{algorithmic}
	\end{algorithm}
	It is important to note that $(k-1)$ transformations are performed on $\ten{a}_k$, computing $\tilde{\ten{a}}_k$, before generating the $k$-th reflector. This can potentially lead to a much larger maximum TT-rank of $\ten{a}_k$ that its initial value. To keep the TT-rank of $\tilde{\ten{a}}_k$ reasonably small, a \texttt{TT-round} is performed before generating the associated Householder TT-vector (see line~\ref{alg:HH:l5}). 
	
	The conclusive part of the TT-Householder transformation algorithm~\ref{alg:HH} generates a new set $\set{Q}\,$ of orthonormal TT-vectors. The $i$-th element of $\set{Q}\;$, $\ten{q}_i$, is obtained by applying the first $i$ Householder reflections in reverse order to $\ten{e}_i$ from the canonical basis $\set{E}$ (see line~\ref{alg:HH:l11} of Algorithm~\ref{alg:HH}). To maintain the maximum TT-rank of $\ten{q}_i$ limited and prevent memory overflow, each $\ten{q}_i$ is rounded to an accuracy $\delta$, as shown in line~\ref{alg:HH:l12} of Algorithm~\ref{alg:HH}. 
	
	\begin{algorithm}[htb]
		\centering
		\caption{$\set{Q}\,, \mat{R}$ = TT-Householder($\set{A}$, $\delta$)}\label{alg:HH} 
		\begin{algorithmic}[1]
			\Input{$\set{A} = \{\ten{a}_1, \dots, \ten{a}_m\}$ a set of TT-vectors, $\delta\in\R_+$ a relative rounding accuracy}
			\Output{$\set{Q} = \{\ten{q}_1, \dots, \ten{q}_m\}$ the set of orthogonal TT-vectors, $\mat{R}$ the upper triangular matrix}
			
			\State let $\set{E} = \{\ten{e}_1, \dots, \ten{e}_m\}$ be the canonical basis of a subspace of dimension $m$ of $\R^{n_1\times\dots\times n_d}$
			\State $\ten{w} = \ten{a}_1$
			\For{$i = 1, \dots, m$} 
			\LineComment construct the $i$-th Householder TT-vector
			\State $\ten{u}_i$, $\mat{R}(:i, i) = \texttt{TTH-vec}(\ten{w}, \set{F}_i, \delta)$ with $\set{F}_i = \{\ten{e}_1, \dots, \ten{e}_i\}$
			\For{$j = i, \dots, m$}		
			\State $\ten{a}_j = \texttt{apply-H-vec}(\ten{a}_j,\ten{u}_i)$\label{alg:HH:l7} \Comment update the $j$-th element of $\set{A}$
			\EndFor
			\If{i < m}
			\LineComment round the TT-vector that will define the successive Householder reflection
			\State $\ten{w} = \texttt{TT-round}(\ten{a}_{i+1}, \delta)$\label{alg:HH:l5}	
			\EndIf
			\EndFor
			
			\For{$i = 1, \dots, m$}	\Comment compute the new orthogonal basis
			\State $\ten{q}_i = \ten{e}_i$
			\For{$j = i, \dots, 1$}
			\State $\ten{q}_i = \texttt{apply-H-vec}(\ten{q}_i,\ten{u}_j)$ for \label{alg:HH:l11} \Comment reflect the $i$-th element of $\set{E}$
			\EndFor
			\State $\ten{q}_i = \texttt{TT-round}(\ten{q}_i, \delta)$\label{alg:HH:l12}
			\EndFor
		\end{algorithmic}
	\end{algorithm}

	The TT-Householder algorithm requires $4m$ \texttt{TT-round} operations, including two for each Householder TT-vector, one for each TT-vector $\tilde{\ten{a}}_k$ after the $(k-1)$-th reflection, and one for each $\ten{q}_k$ orthogonal TT-vector. Therefore, the TT-Householder algorithm is computationally more expensive than all the other orthogonalization methods. Specifically, it is $4$ times more expensive than CGS and MGS, and twice as expensive as CGS2 and MGS2. 
	
	\subsection{Stability comparison}\label{sec2:s4}
	A central issue for the orthogonalization algorithms is the loss of orthogonality, i.e., how much the rounding errors propagate and affect the orthogonality of the computed basis.
	The loss of orthogonality of an orthogonalization scheme applied to the set $\set{A}_m = \{\vec{a}_1, \dots, \vec{a}_m\}$ is defined by the L2-norm of the difference between the identity matrix of size $m$ and the Gram matrix defined by the $m$ vectors generated by the orthogonalization algorithm. We give the definition more formally. Let $\set{Q}\;_m = \{\vec{q}_m, \dots, \vec{q}_m\}$ be a set of $m$ vectors, obtained from an orthogonalization scheme applied to the $m$ vectors of the input set $\set{A}_m$. Let $\vec{q}_i\in\set{Q}\;_m$ be the $i$-th column of the matrix $\mat{Q}_m\in\R^{n\times m}$ for every $i\in\{1, \dots, m\}$, then the Gram matrix associated with the set $\set{Q}\;_m$ is $\mat{Q}_m^\top\mat{Q}_m$. Note that the $(i,j)$ element of $\mat{Q}_m^\top\mat{Q}_m$ is the inner product of $\vec{q}_i$ and $\vec{q}_j$, i.e., $\mat{Q}_m^\top\mat{Q}_m(i,j) = \langle \vec{q}_i,\, \vec{q}_j\rangle$ for every $i,j\in\{1, \dots, m\}$. Then, the loss of orthogonality of the considered algorithm for a basis of size $m$ is equal to 
	\begin{equation}
		\label{eq:LOO}
		||\I_m - \mat{Q}_m^\top\mat{Q}_m||_2.
	\end{equation}
	In the classical matrix framework, an orthogonalization scheme is said to be \emph{numerically stable} if the loss of orthogonality of the basis it computes is of the order of the unit round-off $u$ of the working arithmetic. The following theoretical results, which hold for the six orthogonalization schemes in classical linear algebra, provide a base line for the comparison with the numerical results obtained in the tensor framework, discussed in Section~\ref{sec3}.
	\ignore{Otherwise stated, a stable orthogonalization method propagates orthogonalization errors, measured by the loss of orthogonality and always bounded by the round-off unit\MI{add ref} times a certain real positive constant. \par In the following section, we recall the main theoretical result holding true in the matrix framework for the orthogonalization schemes previously introduced. They constitute a base line for the comparison of the numerical results obtained in the tensor framework, to be discussed in Section~\ref{sec2:s5}.} 
	
	\par In~\cite[Theorem 1]{Giraud2005}, the authors prove that the loss of orthogonality for a basis obtained by CGS, given $\set{A}_m = \{\vec{a}_1, \dots, \vec{a}_m\}$ a set of $m$ vectors, is bounded by a positive constant times the unit round-off $u$ of the working arithmetic, times the squared condition number of the matrix $\mat{A}_m\in\R^{n\times m}$ whose $j$-th column is $\vec{a}_j\in\R^{n}$ for $j\in\{1, \dots, m\}$, i.e.,
	\begin{equation}
	\label{eq:CGS-LOO}
	||\I_m - \mat{Q}_m^\top \mat{Q}_m||_2 \sim \bigO\bigl(u \kappa^2(\mat{A}_m)\bigr)
	\end{equation}
	as long as $\kappa^2(\mat{A}_m)u \ll 1$. 
	In~\cite{Bjorck1967}, an upper bound for the loss of orthogonality of MGS is provided. The loss of orthogonality for a basis of $m$ vectors produced by MGS from $\{\vec{a}_1, \dots, \vec{a}_m\}$ is upper bounded by a constant times the unit round-off $u$ times the condition number of $\mat{A}_m$ as previously defined from $\set{A}_m$ elements, i.e.,
	
	\begin{equation}
	\label{eq:MGS-LOO}
	||\I_m - \mat{Q}_m^\top \mat{Q}_m||_2 \sim \bigO\bigl(u \kappa(\mat{A}_m)\bigr)
	\end{equation}
	as long as $\kappa(\mat{A}_m)u \ll 1$. The authors of~\cite[Theorem 4.1]{Stathopoulos2002}, who proposed the Gram orthogonalization scheme, also estimated an upper bound for the loss of orthogonality of their orthogonalization technique. The loss of orthogonality of a basis of $m$ vectors produced by the Gram scheme from $\{\vec{a}_1, \dots, \vec{a}_m\}$ satisfies the same upper bound as CGS, given in~\eqref{eq:CGS-LOO}. The Householder orthogonalization algorithm is known its stability. The loss of orthogonality of a basis of $m$ vectors produced by Householder transformations from $\{\vec{a}_1, \dots, \vec{a}_m\}$ is bounded by a constant times the round-off unit, i.e.,
	\begin{equation}
	\label{eq:HH-LOO}
	||\I_m - \mat{Q}_m^\top \mat{Q}_m||_2 \sim \bigO\bigl(u\bigr)
	\end{equation}
	as proven in~\cite{Wilkinson1988}. When introducing a further orthogonalization step in the classical and modified Gram-Schmidt, defining CGS2 and MGS2, their loss of orthogonality improves considerably, reaching Householder quality. As proven in~\cite{Giraud2005, Smoktunowicz2006}, the loss of orthogonality of CGS2 and MGS2 satisfies the bound given in Equation~\eqref{eq:HH-LOO}, under the hypothesis $\kappa^2(A_m)u \ll 1$ for CGS2, while it holds for MGS2 if $\kappa(A_m)u \ll 1$. Table~\ref{tab:orth} presents a summary of all the loss of orthogonality bounds.
	
	\section{Numerical tensor experiments}\label{sec3}
	Sections~\ref{sec2:s1}~-~\ref{sec2:s3} describe four orthogonalization methods that produce an orthonormal basis of TT-vectors, given a set of TT-vectors and a rounding accuracy $\delta$. This section analyzes two sets of results obtained from the orthogonalization schemes, highlighting similarities and differences with the known theoretical results in matrix computation. In all the experiments, the input set of TT-vectors $\set{A}_m = \{\ten{a}_1, \dots, \ten{a}_m\}$ is generated using a Krylov process. Starting with a TT-vector of ones, $\ten{x}_1\in\R^{n_1\times \dots \times n_d}$, we iteratively compute $\ten{x}_{j+1} = -\bm{\Delta}_d\ten{a}_j$ where $\bm{\Delta}_d$ is the TT-matrix representing the discretization of the Laplacian operator of order $d$ with Dirichlet boundary conditions, see \cite{Kazeev2012}, and $\ten{a}_j$ is the normalized output of the \texttt{TT-round} algorithm applied to $\ten{x}_j$\MI{with the accuracy replaced by a maximum TT-rank equal to $1$, see~\cite{Oseledets2011TT} for further details}. As a result, $\ten{a}_j$, the $j$-th element of $\set{A}_m$, has a TT-rank of $1$ for every $j\in\{1, \dots, m\}$. This TT-rank constraint facilitates the analysis of the memory requirement. The elements of $\set{A}$ are generated as a sequence of $m$ normalized (and rounded) Krylov TT-vectors. Therefore, if the first $k$ of them are vectorized and arranged as columns of the matrix $\mat{A}_k$, the condition number $\kappa(A_k)$ grows for $k\in\{1, \dots, m\}$. The $\set{A}_k$ denotes the subset of $\set{A}_m$ defined by its first $k$ TT-vectors.
	The two experiments differ in the problem dimension, with the first having a dimension of $3$ and the second having dimension $6$, but they have the same mode size $n = 15$. Further details are provided in the following sections.
	
	\subsection{Numerical loss of orthogonality}\label{sec3:s1}
	This section examines the numerical results of two sets of experiments from the perspective of the loss of orthogonality. The purpose is to highlight the similarities with the classical matrix orthogonalization methods. Specifically, we investigate the loss of orthogonality $\norm{\I_k - \mat{Q}_k^\top \mat{Q}_k}_2$,
	where the $(i,j)$ element of $\mat{Q}_k^\top \mat{Q}_k$ is computed by the inner product of the $i$-th and $j$-th TT-vector of the orthogonal basis, i.e., 
	\begin{equation*}
		\mat{Q}_k^\top \mat{Q}_k(i,j) = \langle \ten{q}_i, \ten{q}_j\rangle
	\end{equation*}
	for $\ten{q}_i, \ten{q}_j\in\set{Q}\,$ for $i,j\in\{1, \dots, k\}$ for $k\in\{1, \dots, m\}$, where $\set{Q}\,$ denotes the set of TT-vectors produced by the considered orthogonalization kernel.
	\subsection{Order-$3$ experiments}
	\begin{figure}[htb]
		\centering
		\subfloat[$\delta = 10^{-3}$]{\includegraphics[scale = 0.5,width=0.3\textwidth]{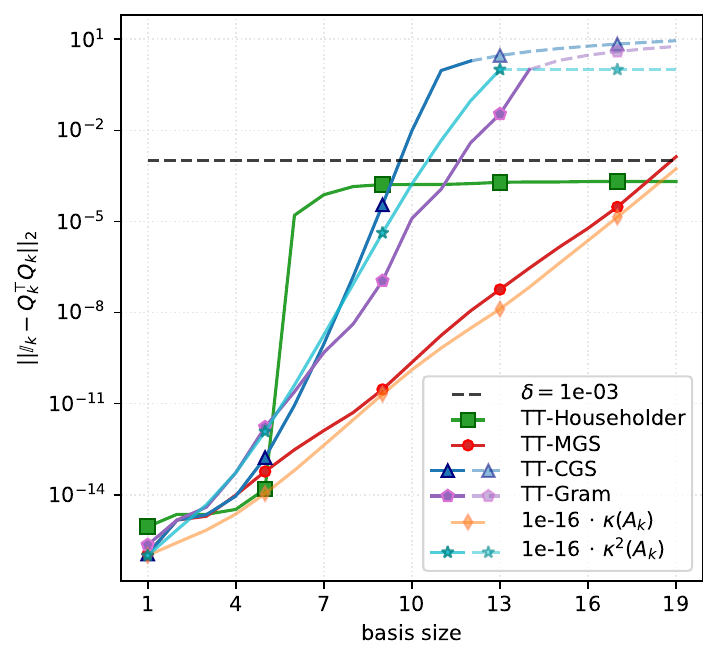}\label{fig:LOO:CN:3}}
		\hfill
		\subfloat[$\delta = 10^{-5}$]{\includegraphics[scale = 0.5,width=0.3\textwidth]{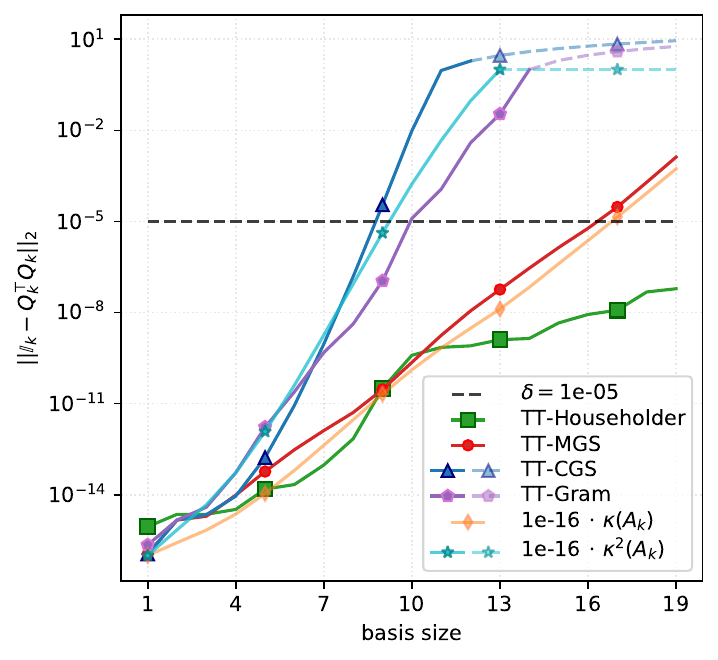}\label{fig:LOO:CN:5}}
		\hfill
		\subfloat[$\delta = 10^{-8}$]{\includegraphics[scale = 0.5,width=0.3\textwidth]{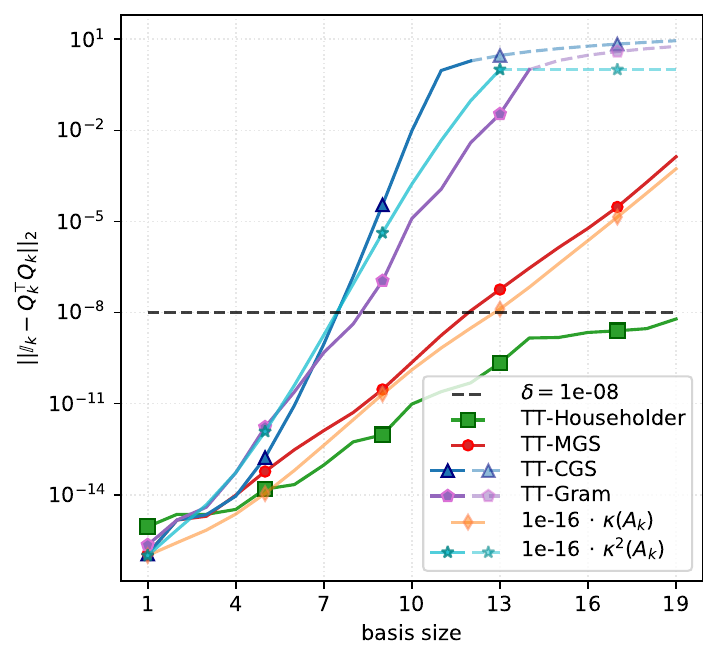}\label{fig:LOO:CN:8}}		\vskip0.1\baselineskip	
		\centering\textit{Loss of orthogonality with classical algorithms}\\
		\vskip2\baselineskip
		\subfloat[$\delta = 10^{-3}$]{\includegraphics[scale = 0.5,width=0.3\textwidth]{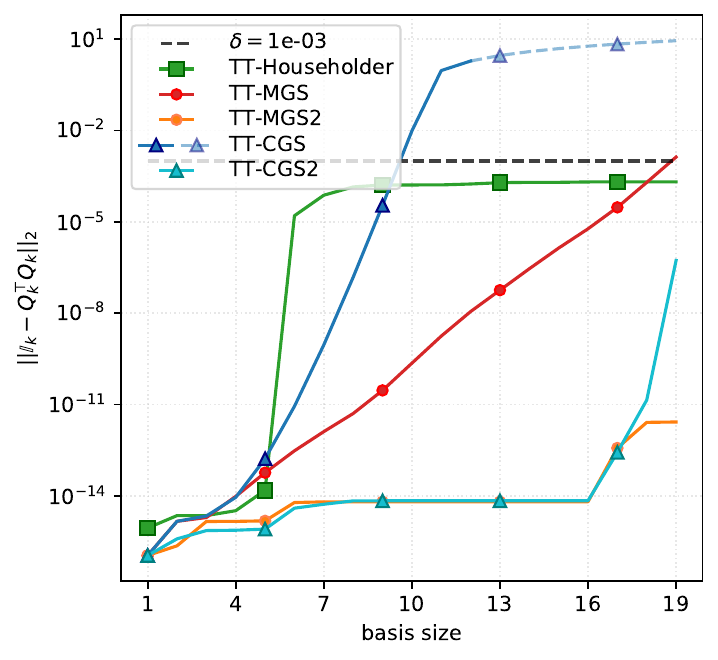}\label{fig:LOO:O2:3}}
		\hfill
		\subfloat[$\delta = 10^{-5}$]{\includegraphics[scale = 0.5,width=0.3\textwidth]{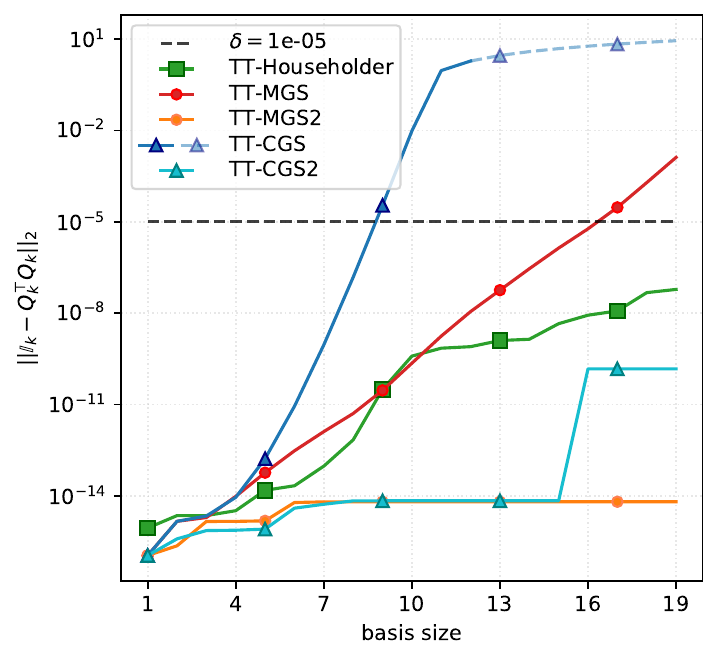}\label{fig:LOO:O2:5}}
		\hfill
		\subfloat[$\delta = 10^{-8}$]{\includegraphics[scale = 0.5,width=0.3\textwidth]{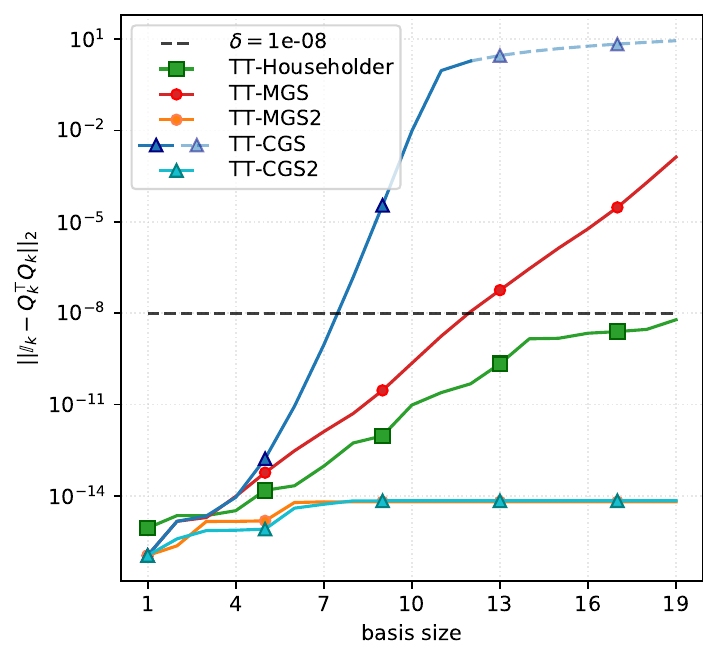}\label{fig:LOO:O2:8}}
		\vskip0.1\baselineskip	
		\centering\textit{Loss of orthogonality with re-orthogonalization comparison}\\
		
		\caption{Loss of orthogonality and condition number for $m = 20$ TT-vectors of order $d = 3$ and mode size $n = 15$. The curves get dashed and partially transparent when they get greater than~$1$.}
		\label{fig:LOO:CN}
	\end{figure} 
	
	In the the first experiment, we set the order $d = 3$, the size mode $n_i = 15$ for $i\in\{1,2,3\}$, and the input number of TT-vectors $m = 20$. Figure~\ref{fig:LOO:CN} reports the results of this first group of experiments for the six different schemes and three different rounding accuracy values $\delta\in\{10^{-3}, 10^{-5}, 10^{-8}\}$. The use of a low dimensional problem allows us to convert each TT-vector $\ten{a}_j$ into a dense format and vectorize it. These vectors are stored as the $j$-th column of $\mat{A}_m\in\R^{n^3\times m}$. Consequently, we can estimate the condition number of $\mat{A}_k\in\R^{n^3\times k}$, which is the submatrix of $\mat{A}_m$ formed by its first $k$ columns. Figure~\ref{fig:LOO:CN} displays the loss of orthogonality with colored continuous curves, and the constant rounding accuracy $\delta$ with a dashed black line in all plots. The condition number $\kappa(\mat{A}_k)$ and its squared value scaled by $u\approx 10^{-16}$ are also shown with colored continuous lines, as long as they are smaller than $1$. 
	For ease of comparison with the slope of the loss of orthogonality of TT-MGS and TT-CGS, the condition number curves are scaled. All the curves are dependent on the basis size $k$. As previously mentioned, the TT-vectors are generated as a sequence of normalized Krylov TT-vectors. As the value of $k$ increases, the elements of $\set{A}_k$ become more linearly dependent, resulting in an increase in associated condition number $\kappa(\mat{A}_k)$.
	
	To aid interpretation, we present three plots in Figure~\ref{fig:LOO:CN:3},~\ref{fig:LOO:CN:5} and~\ref{fig:LOO:CN:8}, which show the loss of orthogonality of standard methods in tensor format: TT-Householder, TT-MGS, TT-CGS, TT-Gram. Figures~\ref{fig:LOO:O2:3},~\ref{fig:LOO:O2:5} and~\ref{fig:LOO:O2:8} show the loss of orthogonality of re-orthogonalization methods: TT-MGS2 and TT-CGS2, compared to TT-Householder, TT-MGS and TT-CGS.
	All six plots in Figure~\ref{fig:LOO:CN} exhibit similar behaviors. The loss of orthogonality of the TT-Householder method in green stagnates around the rounding accuracy $\delta$, as shown in Figure~\ref{fig:LOO:CN:3}. This is consistent with the matrix theoretical expectation, stated in Equation~\eqref{eq:HH-LOO}, where the unit round-off $u$ is replaced by the \texttt{TT-round} accuracy $\delta$.	
	The loss of orthogonality of TT-MGS method in red grows with the same slope as the condition number $\kappa(\mat{A}_k)$ in dashed green, matching the matrix upper bound stated in~\eqref{eq:MGS-LOO}. Finally, both the TT-CGS and and TT-Gram loss of orthogonality curves cross the rounding accuracy dashed line faster than TT-MGS. This curve follows the the squared condition number $\kappa^2(\mat{A}_k)$, as long as $\kappa^2(\mat{A}_k) < 10^2$. The loss of orthogonality curves for TT-CGS and TT-Gram stagnates below $10^{2}$ for $k > 10$ approximately, while $\kappa^2(\mat{A}_k)$ continues to grow. Upon analyzing the second line plots, it is evident that Figure~\ref{fig:LOO:O2:3},~\ref{fig:LOO:O2:5} and~\ref{fig:LOO:O2:8} demonstrate a significant improvement in the the loss of orthogonality when a second re-orthogonalization loop is introduced. It is noteworthy that the loss of orthogonality of TT-CGS2 is close to the machine precision, approximately $10^{-14}$ for $\delta\in\{10^{-3}, 10^{-5}\}$, increasing after for $k \ge 15$. For $\delta = 10^{-8}$, it remains around $10^{-14}$. Therefore, as long as the elements of $\set{A}_k$ are not highly collinear, TT-CGS2 outperforms TT-CGS, TT-MGS and TT-Householder, but not TT-MGS2. For the rounding accuracy $\delta\in\{10^{-5}, 10^{-8}\}$, the loss of orthogonality of TT-MGS2 remains around $10^{-14}$, while for $\delta = 10^{-3}$ the loss of orthogonality jumps from $10^{-14}$ to $10^{-11}$, where it appears to remain constant, when $k > 16$. Overall ,TT-MGS2 is the best performing algorithm among all the others. The results for TT-CGS2 and TT-MGS2 are consistent with the matrix theory presented in Section~\ref{sec2:s4}. It is hypothesized that the jumps occur when the condition number $\kappa(\mat{A}_k)$, or its square, multiplied by the rounding accuracy is no longer sufficiently smaller than $1$.
	\subsection{Order-$6$ experiments}
	\begin{figure}[htb]
		\centering
		\subfloat[$\delta = 10^{-3}$]{\includegraphics[scale = 0.5,width=0.3\textwidth]{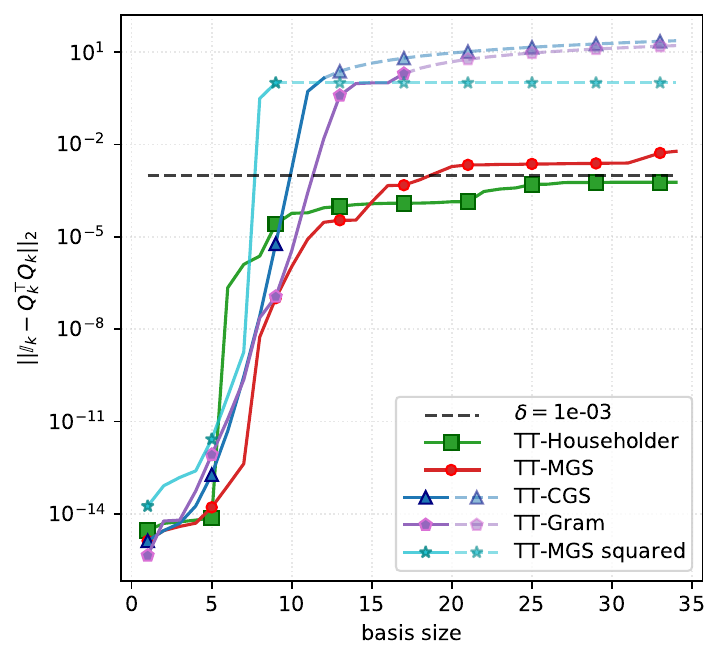}\label{fig:LOO:3}}
		\hfill
		\subfloat[$\delta = 10^{-5}$]{\includegraphics[scale = 0.5,width=0.3\textwidth]{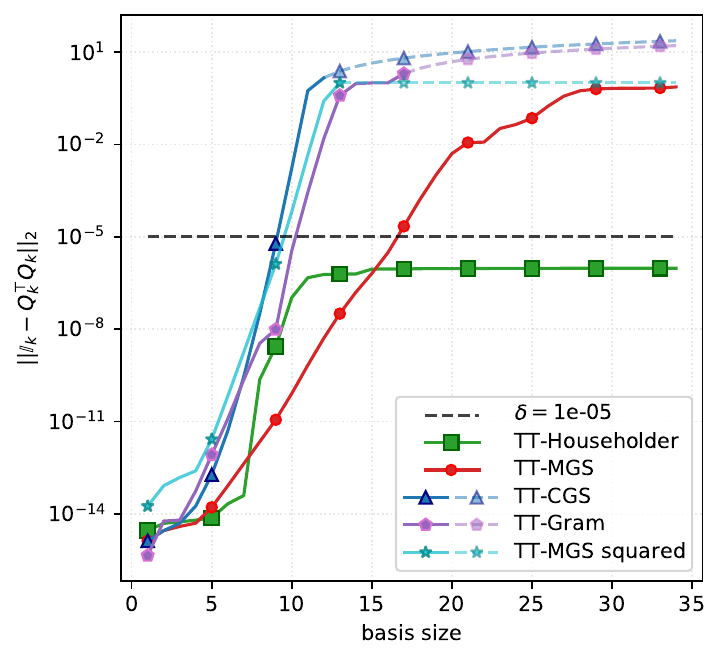}\label{fig:LOO:5}}
		\hfill
		\subfloat[$\delta = 10^{-8}$]{\includegraphics[scale = 0.5,width=0.3\textwidth]{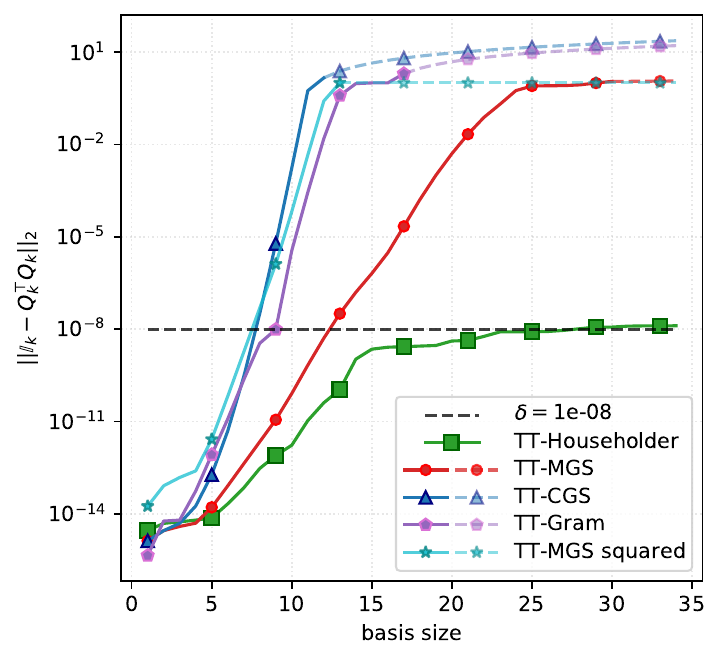}\label{fig:LOO:8}}\\
		\vskip0.1\baselineskip	
		\centering\textit{Loss of orthogonality with classical algorithms}\\
		
		\vskip2\baselineskip
		\subfloat[$\delta = 10^{-3}$]{\includegraphics[scale = 0.5,width=0.3\textwidth]{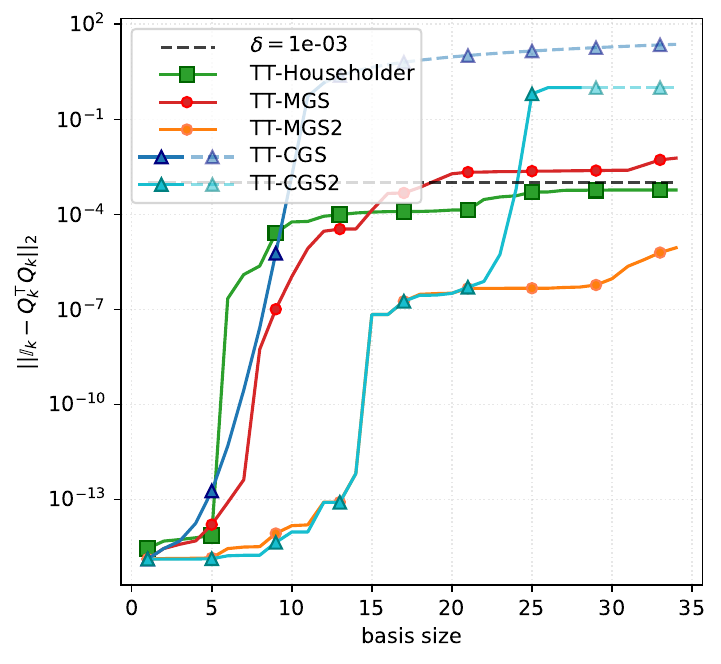}\label{fig:LOO:O26d:3}}
		\hfill
		\subfloat[$\delta = 10^{-5}$]{\includegraphics[scale = 0.5,width=0.3\textwidth]{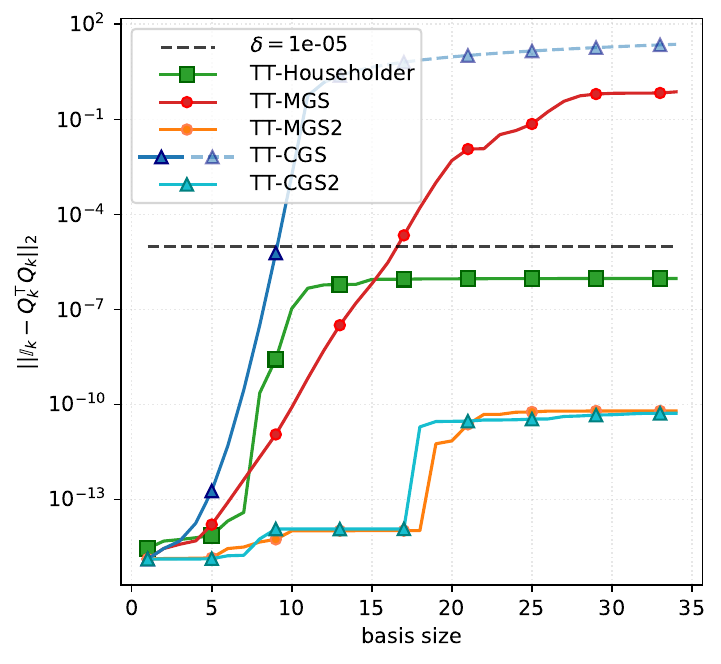}\label{fig:LOO:O26d:5}}
		\hfill
		\subfloat[$\delta = 10^{-8}$]{\includegraphics[scale = 0.5,width=0.3\textwidth]{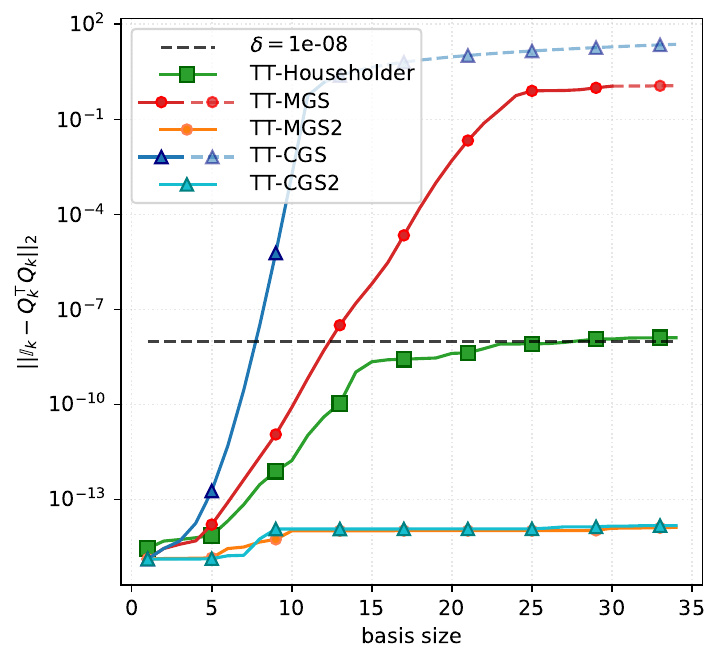}\label{fig:LOO:O26d:8}}
		\vskip0.1\baselineskip	
		\centering\textit{Loss of orthogonality with re-orthogonalization comparison}\\
		\vskip\baselineskip

		\caption{Loss of orthogonality for $m = 35$ TT-vectors of order $d = 6$ and mode size $n = 15$. The curves get dashed and partially transparent when they get greater than $1$.} 
		\label{fig:LOO:6d}
	\end{figure}
	
	To further validate our results and to study their applicability to large-scale problems, we introduce a second experimental framework. We set the problem order to $d = 6$ with size mode $n_i = 15$ for $i\in\{1, \dots, 6\}$, and generate $m = 35$ TT-vectors, defining the set $\set{A} = \{\ten{a}_1, \dots, \ten{a}_{35}\}$. For the rounding accuracy values $\delta\in\{10^{-3}, 10^{-5}, 10^{-8}\}$, we compute the loss of orthogonality for the four orthogonalization schemes, presented in Section~\ref{sec2:s1}~-~\ref{sec2:s3}. Figure~\ref{fig:LOO:6d} displays the loss of orthogonality of these experiments. Due to the problem order $d = 6$ and size $n = 15$, the curve of the condition number of the matrix $\mat{A}_k\in\R^{n^6\times k}$ is not included. 
	To compensate for the absence of the condition number curve, we display the square of the TT-MGS loss of orthogonality values with a dashed line. This line should exhibit the same slope as the CGS loss of orthogonality (and consequently of the squared condition number $\kappa(A_k)$), if the matrix theory extends to the TT-framework. Similarly to the previous case, the first line of plots displays standard orthogonalization algorithms in TT-format. The second line shows results from methods with re-orthogonalization. Figures~\ref{fig:LOO:3}, ~\ref{fig:LOO:5} and~\ref{fig:LOO:8} demonstrate that the TT-Householder orthogonalization algorithm produces a basis with a loss of orthogonality that stagnates around the rounding accuracy $\delta$ for every value in $\{10^{-3}, 10^{-5}, 10^{-8}\}$ after the basis size exceeds approximately $10$. This supports the intuition that the bound expressed in Equation~\eqref{eq:HH-LOO} still holds true in the tensor framework, with the unit round-off $u$ replaced by the \texttt{TT-round} accuracy $\delta$. In Figures~\ref{fig:LOO:3} and~\ref{fig:LOO:5}, when the basis size is smaller than about $15$, the TT-MGS loss of orthogonality is smaller than the Householder one. However, when the basis includes more then $15$ TT-vectors, the relation reverses. For more accurate computation, specifically for $\delta = 10^{-8}$, the Householder loss of orthogonality outperforms the TT-MGS loss of orthogonality when the basis size is around $5$. Note that the TT-MGS loss of orthogonality increases linearly and then stabilizes at different level for each rounding accuracy. It reaches $10^{-2}$ for $\delta = 10^{-3}$ and $1$ for $\delta\in\{10^{-5}, 10^{-8}\}$. The TT-CGS and TT-Gram loss of orthogonality curves reaches stabilization almost immediately when the basis size is greater than $10$ for all the rounding accuracy values, as shown in Figure~\ref{fig:LOO:6d}. This is likely due to the poor condition number of the input, rendering the assumptions of matrix theory invalid. Specifically, the condition number multiplied by the rounding accuracy smaller than $1$. Furthermore, Figures~\ref{fig:LOO:5} and~\ref{fig:LOO:8} demonstrate that the loss of orthogonality of TT-CGS and TT-Gram follows the square of loss of orthogonality of the TT-MGS. This supports the idea that even in the TT-format, the loss of orthogonality of TT-MGS increases as the condition number, while the loss of orthogonality of TT-Gram and TT-CGS increases as the squared condition number grows. Additionally, Figures~\ref{fig:LOO:O26d:3}, ~\ref{fig:LOO:O26d:5} and~\ref{fig:LOO:O26d:8} display the loss of orthogonality of TT-MGS2 and TT-CGS2 compared to the previously analyzed results of TT-Householder, TT-MGS and TT-CGS. TT-MGS2 outperforms all the other methods for all considered rounding accuracies. Its loss of orthogonality stagnates around $10^{-5}$ for $\delta = 10^{-3}$, around $10^{-10}$ for $\delta = 10^{-5}$ and around $10^{-13}$ for $\delta = 10^{-8}$. In Figures~\ref{fig:LOO:O26d:3}-~\ref{fig:LOO:O26d:8}, the TT-MGS2 curve shows a larger jump for greater values of $\delta$, for $15 \le k \le 20$ with $\delta\in\{10^{-3}, 10^{-5}\}$ and for $k \sim 10$ with $\delta = 10^{-8}$. TT-CGS2 outperforms the TT-Householder, TT-CGS and TT-MGS, similarly to TT-MGS2, as long as the TT-vectors are not highly collinear (i.e., for $k < 20$ when $\delta = 10^{-3}$) for $\delta\in\{10^{-5}, 10^{-8}\}$. These results support the conclusion made for the $d = 3$ experiments, that the bounds for the loss of orthogonality of TT-CGS2 and TT-MGS2 established in the classical matrix framework still hold, possibly under revised assumptions.
	
	\subsection{Memory usage estimation}\label{sec3:s2}
	This section aims to analyze the effects of the orthogonalization process on the TT-rank and memory requirement based on experimental results. The growth of the TT-ranks, of the compression ratio (defined in Equation~\eqref{eq:comp_ratio}), and the compression gain (cf. Equation~\eqref{eq:gain_ratio}) curve of the orthogonal basis are investigated in the second set of experiments with $d = 6$, $n_i = 15$, and $m = 35$. This setting can be considered a large-scale one, and we use as rounding accuracy values $\delta\in\{10^{-3}, 10^{-5}, 10^{-8}\}$ 
	\subsubsection{Householder transformation}
		\begin{figure}[p]
		\centering
		\subfloat[$\delta = 10^{-3}$]{\includegraphics[scale = 0.5,width=0.3\textwidth]{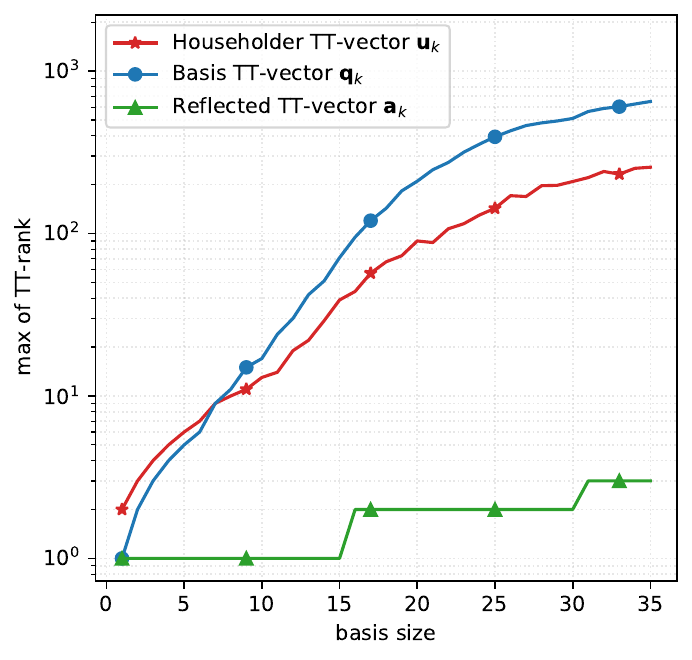}\label{fig:mem:HH:rk:3}}
		\hfill
		\subfloat[$\delta = 10^{-5}$]{\includegraphics[scale = 0.5,width=0.3\textwidth]{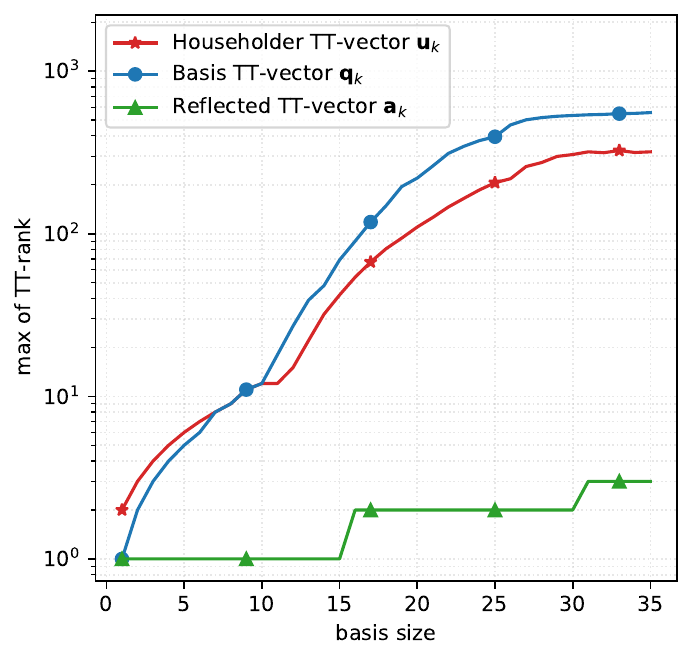}\label{fig:mem:HH:rk:5}}
		\hfill
		\subfloat[$\delta = 10^{-8}$]{\includegraphics[scale = 0.5,width=0.3\textwidth]{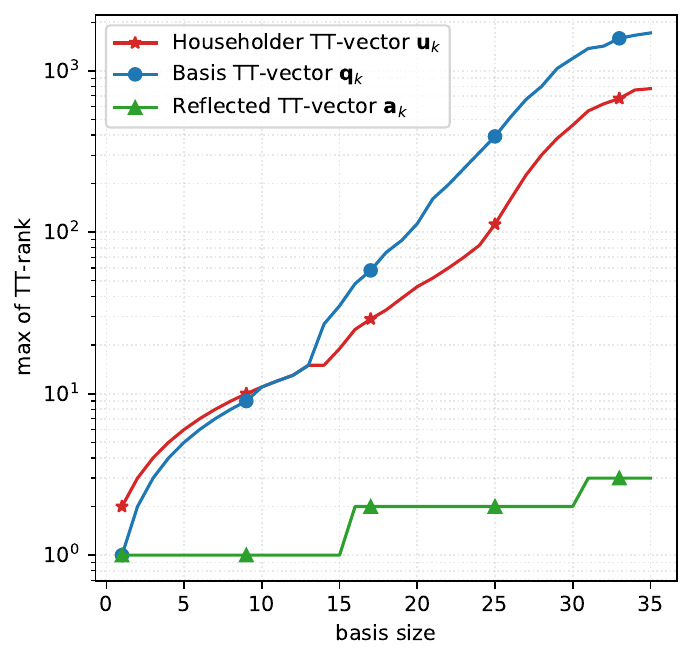}\label{fig:mem:HH:rk:8}}
		\vskip0.1\baselineskip
		\centering\textit{Maximal TT-rank for the TT-vectors in Householder algorithm}\\
		\vskip2\baselineskip
		
		\subfloat[$\delta = 10^{-3}$]{\includegraphics[scale = 0.5,width=0.3\textwidth]{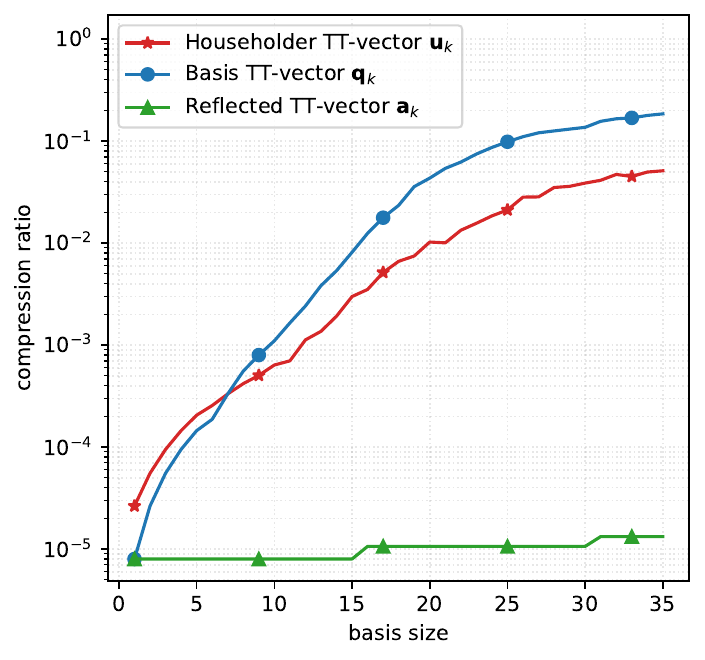}\label{fig:mem:HH:cr:3}}
		\hfill
		\subfloat[$\delta = 10^{-5}$]{\includegraphics[scale = 0.5,width=0.3\textwidth]{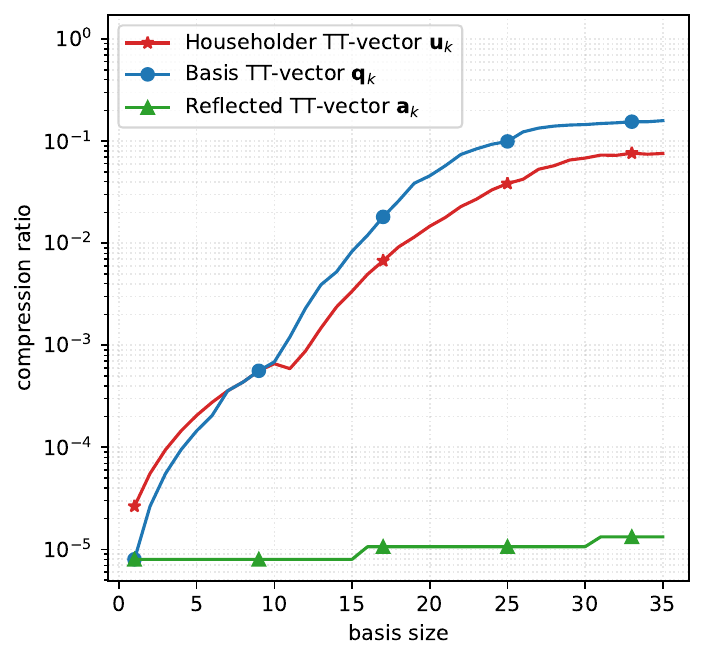}\label{fig:mem:HH:cr:5}}
		\hfill
		\subfloat[$\delta = 10^{-8}$]{\includegraphics[scale = 0.5,width=0.3\textwidth]{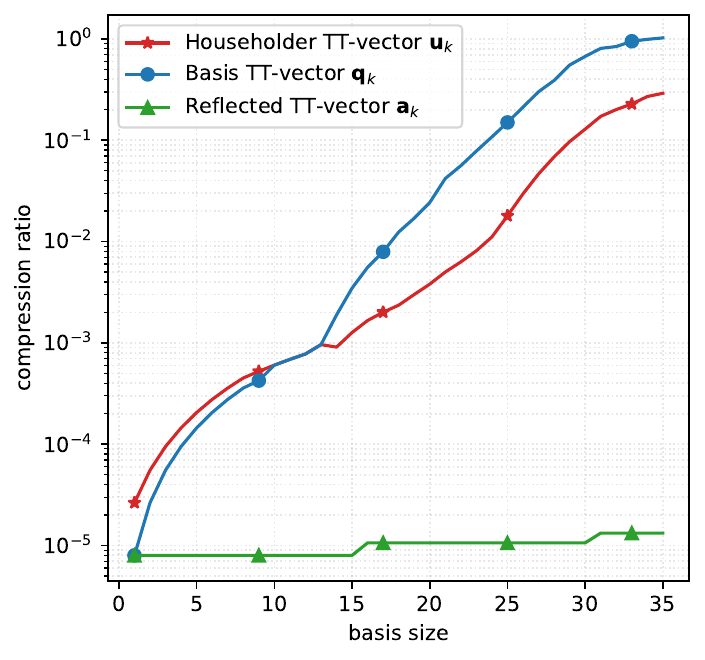}\label{fig:mem:HH:cr:8}}	
		\vskip0.1\baselineskip	
		\centering\textit{Compression ratio for the TT-vectors in Householder algorithm}\\
		
		\vskip2\baselineskip
		
		\subfloat[$\delta = 10^{-3}$]{\includegraphics[scale = 0.5,width=0.3\textwidth]{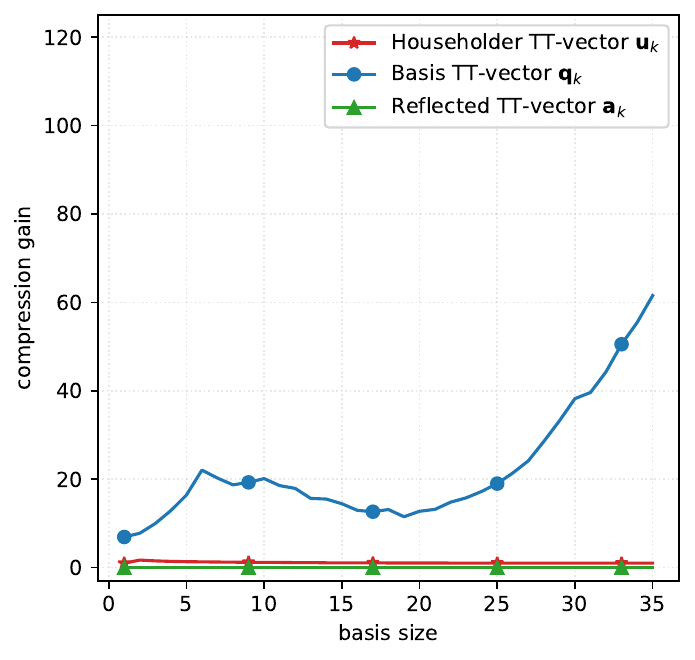}\label{fig:mem:HH:g:3}}
		\hfill
		\subfloat[$\delta = 10^{-5}$]{\includegraphics[scale = 0.5,width=0.3\textwidth]{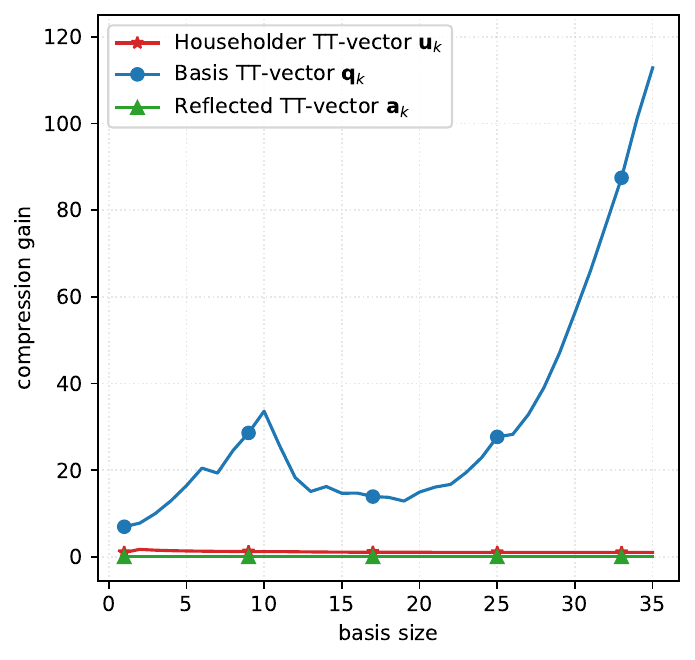}\label{fig:mem:HH:g:5}}
		\hfill
		\subfloat[$\delta = 10^{-8}$]{\includegraphics[scale = 0.5,width=0.3\textwidth]{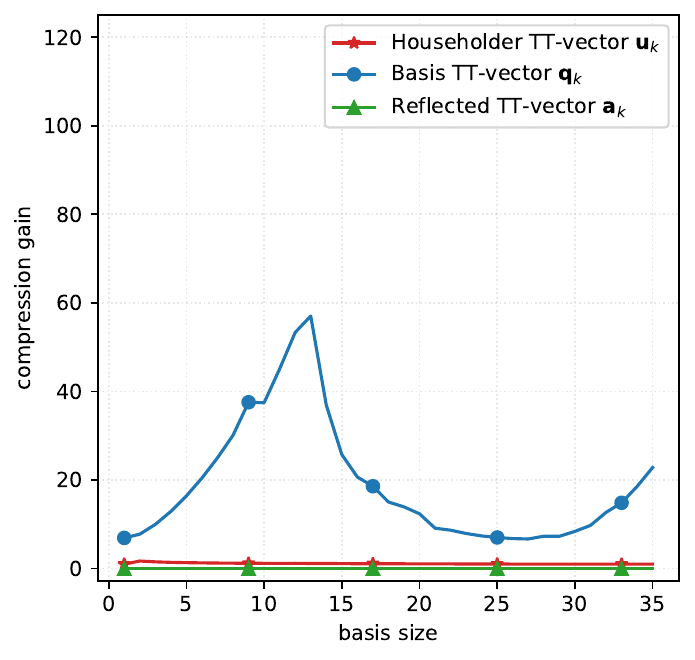}\label{fig:mem:HH:g:8}}	
		\vskip0.1\baselineskip	
		\centering\textit{Compression gain for the TT-vectors in Householder algorithm}\\
		\vskip\baselineskip
		
		\caption{Householder memory requirement for $m = 35$ TT-vectors of order $d = 6$ and mode size $n = 15$.} 
		\label{fig:mem:HH:6d}
	\end{figure} 
	The Householder algorithm~\ref{alg:HH} applies the \texttt{TT-round} to three sets of TT-vectors: the Householder TT-vector $\ten{u}_k$, the TT-vector $\ten{a}_k$ (to which $k$ Householder transformations are applied) and the orthogonal TT-vector $\ten{q}_k$ (obtained from the canonical basis TT-vectors with $i$ successive Householder transformations). It is important to study the evolution of the maximum TT-rank, of the compression ratio, and gain for each of these three groups of TT-vector. Figure~\ref{fig:mem:HH:6d} displays the maximum TT-rank, the compression ratio, and the compression gain of $\ten{u}_k$, $\ten{a}_k$ (after the $k$-th reflection), and $\ten{q}_k$ for every $k\in\{1, \dots, 35\}$ and all values of the rounding accuracy $\delta$. The maximum TT-rank and the compression ratio of $\ten{u}_k$, $\ten{a}_k$ and $\ten{q}_k$ increase with increasing basis sizes, $k$, as expected, due to the growing number of terms in their computation. Notably, the maximum TT-rank of $\ten{q}_k$ exceeds that of $\ten{u}_k$ for a basis size greater than $10$. This property is important because in most practical vector computations, only the Householder TT-vectors $\ten{u}_k$ are stored, and the orthogonal basis TT-vector $\ten{q}_k$ is usually not explicitly formed. \\ Figures~\ref{fig:mem:HH:rk:3},~\ref{fig:mem:HH:rk:5} and~\ref{fig:mem:HH:rk:8} show that the maximum TT-rank of $\ten{a}_k$ after the $k$-th Householder reflection is extremely low, especially when compared to those of $\ten{q}_k$ and $\ten{u}_k$. Furthermore, the maximum TT-rank of $\ten{a}_k$ increases every time the index $k$ is a multiple of the mode size $n = 15$, as shown in Figure~\ref{fig:mem:HH:rk:3} and~\ref{fig:mem:HH:rk:5}. This pattern is also observed in Figure~\ref{fig:mem:HH:rk:8} for $\delta = 10^{-8}$, although it is less consistent. We attempted to explore this phenomenon theoretically, but we were unable to provide a clear and convincing explanation. \\
	The compression ratio closely follows the same trend as the maximum TT-rank, but it allows us to monitor memory growth as a percentage. Figures~\ref{fig:mem:HH:cr:3},~\ref{fig:mem:HH:cr:5} and~\ref{fig:mem:HH:cr:8} show a compression ratio smaller than $1$ for all rounding accuracy $\delta$ values, indicating that the TT-format is still effective in reducing the memory usage in all these experiments. Additionally, the compression ratio of $\ten{a}_k$ remains around $10^{-5}$ for all three values of $\delta$. Figures~\ref{fig:mem:HH:cr:3} and~\ref{fig:mem:HH:cr:5} indicate that the compression ratio of $\ten{q}_k$ plateaus at approximately $10^{-1}$, while that of $\ten{u}_k$ plateaus at around $10^{-2}$. In Figure~\ref{fig:mem:HH:cr:8}, it is unclear whether the compression ratio of $\ten{q}_k$ and $\ten{u}_k$ plateaus at $1$ and $10^{-1}$ respectively. In Figures~\ref{fig:mem:HH:g:3},~\ref{fig:mem:HH:g:8} and~\ref{fig:mem:HH:g:5}, the gain curves of $\ten{u}_k$, $\ten{q}_k$ and $\ten{a}_k$ exhibit same behavior. Specifically, we examine the gain of compressing the $k$-th Householder TT-vector $\ten{u}_k$ and the input TT-vector $\ten{a}_k$ (after the $k$-th reflection). These curves are almost constant and low during the iterations. The several compression steps of the associated TT-vectors during the previous iterations are likely the cause. In contrast, the compression gain for $\ten{q}_k$ is significantly larger. The compression gain curve of $\ten{q}_k$ increases during the first $10$ iterations for all rounding accuracies, decreases slightly after, and seems to raise again. It is worth noting that the compression gain curve of the Householder basis reaches its highest value at around $110$ when $\delta = 10^{-5}$. This indicates that after the compression, slightly less than $1\%$ of the memory used to store the same tensor in TT-format before the compression is required.

	\subsubsection{Orthogonalization kernel comparison}
\begin{figure}[p]
		\centering
		\subfloat[ $\delta = 10^{-3}$]{\includegraphics[scale = 0.5,width=0.3\textwidth]{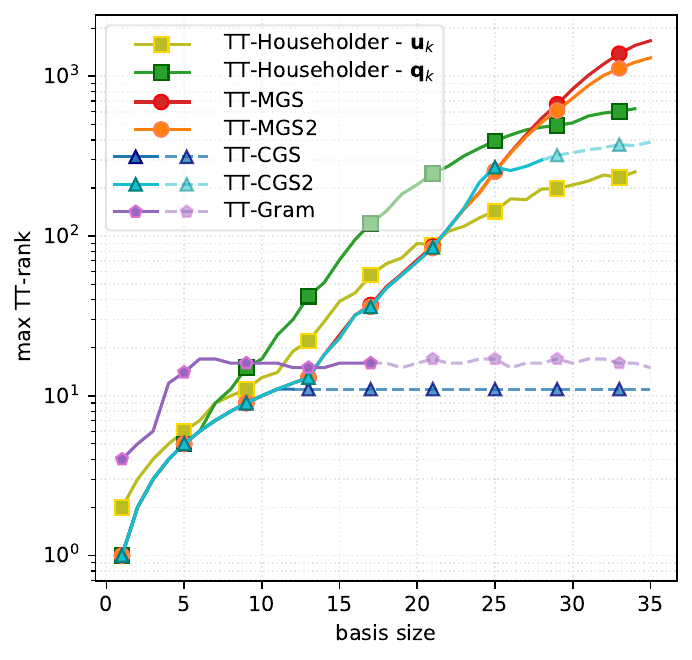}\label{fig:mem:6d:rk:3}}
		\hfill
		\subfloat[$\delta = 10^{-5}$]{\includegraphics[scale = 0.5,width=0.3\textwidth]{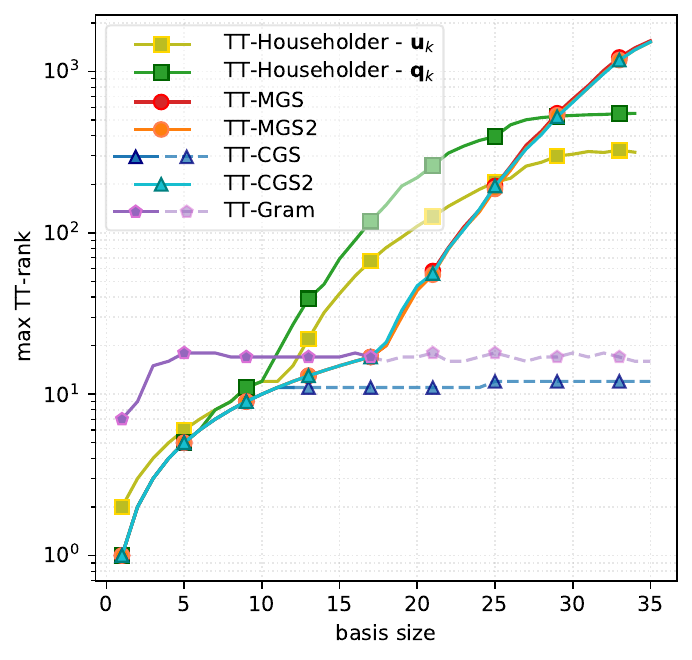}\label{fig:mem:6d:rk:5}}
		\hfill
		\subfloat[$\delta = 10^{-8}$]{\includegraphics[scale = 0.5,width=0.3\textwidth]{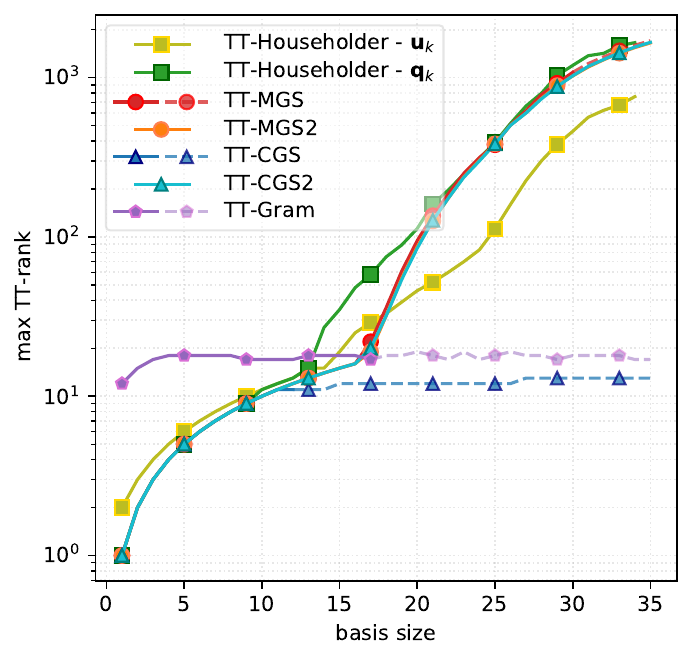}\label{fig:mem:6d:rk:8}}
		\vskip0.1\baselineskip
		\centering\textit{Maximal TT-rank for the orthogonal basis}\\
		\vskip2\baselineskip
		
		\subfloat[$\delta = 10^{-3}$]{\includegraphics[scale = 0.5,width=0.3\textwidth]{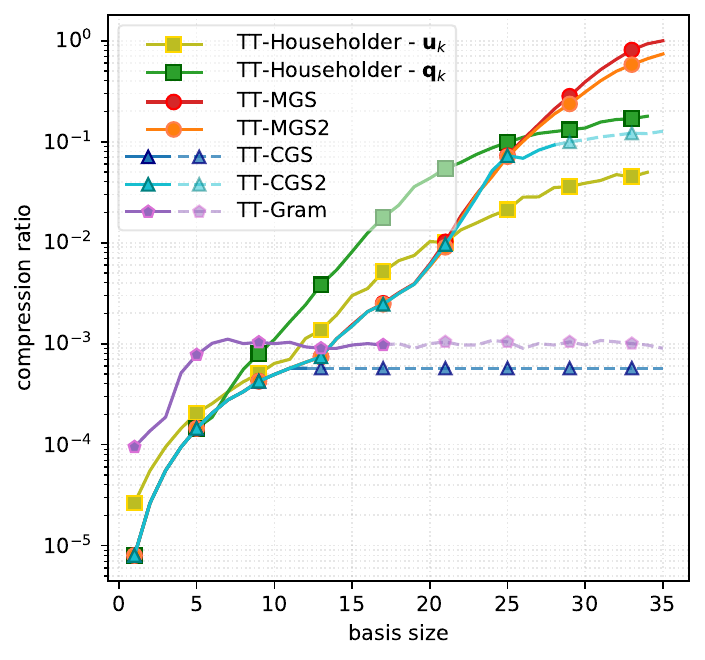}\label{fig:mem:6d:cr:3}}
		\hfill
		\subfloat[$\delta = 10^{-5}$]{\includegraphics[scale = 0.5,width=0.3\textwidth]{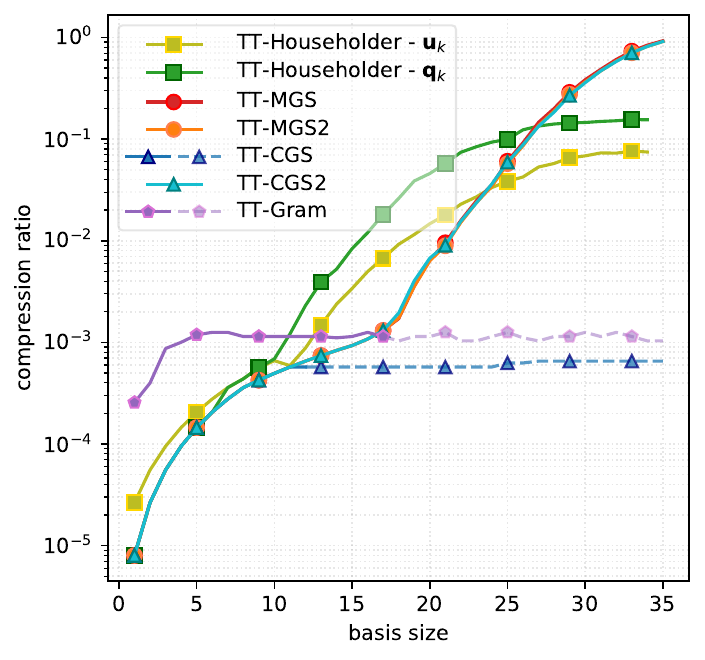}\label{fig:mem:6d:cr:5}}
		\hfill
		\subfloat[$\delta = 10^{-8}$]{\includegraphics[scale = 0.5,width=0.3\textwidth]{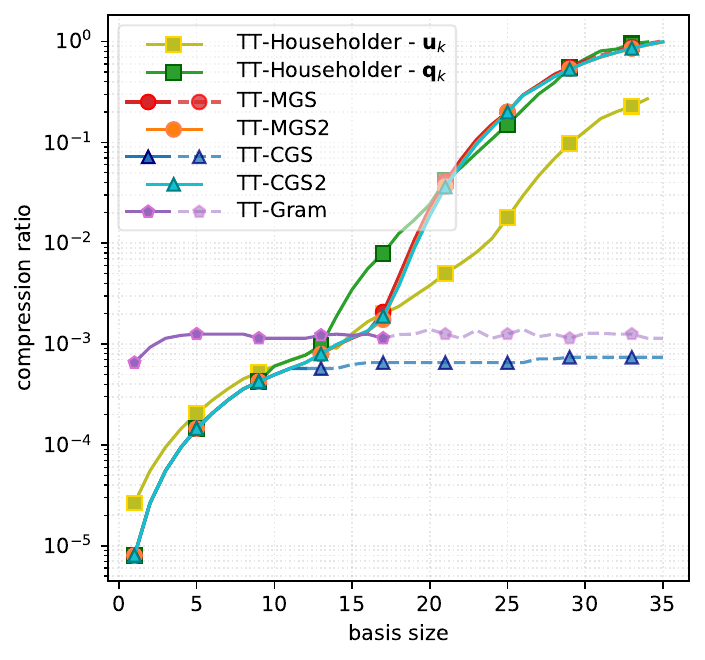}\label{fig:mem:6d:cr:8}}
		\vskip0.1\baselineskip	
		\centering\textit{Compression ratio for the orthogonal basis}\\
		
		\vskip2\baselineskip
		\subfloat[$\delta = 10^{-3}$]{\includegraphics[scale = 0.5,width=0.3\textwidth]{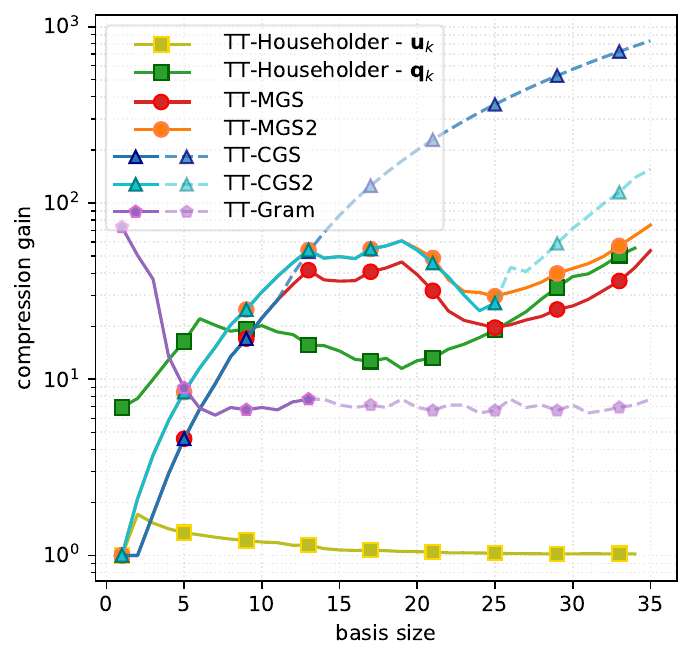}\label{fig:mem:6d:g:3}}
		\hfill
		\subfloat[$\delta = 10^{-5}$]{\includegraphics[scale = 0.5,width=0.3\textwidth]{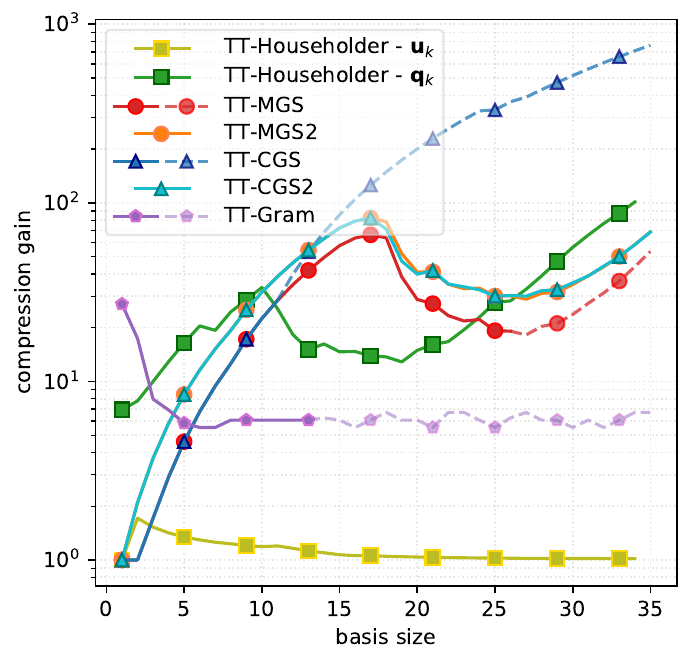}\label{fig:mem:6d:g:5}}
		\hfill
		\subfloat[$\delta = 10^{-8}$]{\includegraphics[scale = 0.5,width=0.3\textwidth]{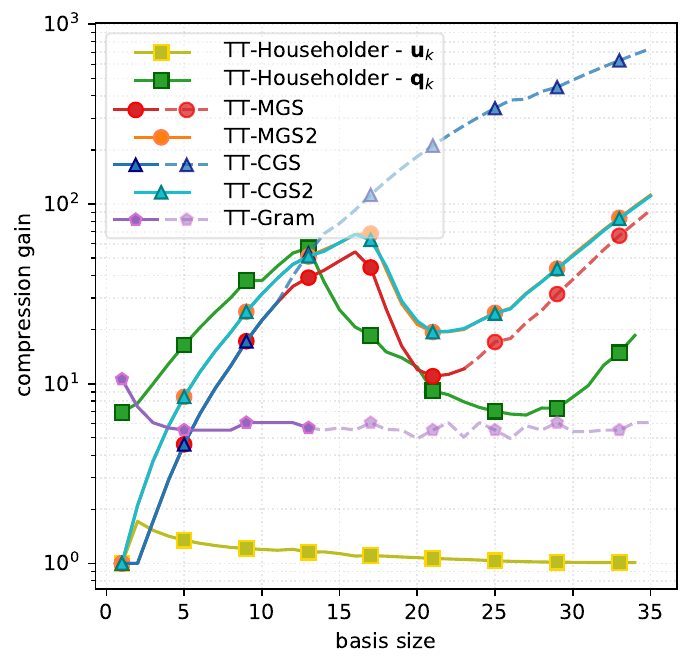}\label{fig:mem:6d:g:8}}
		\vskip0.1\baselineskip	
		\centering\textit{Compression gain for the orthogonal basis}\\
		\vskip\baselineskip
		
		\caption{Comparison of the orthogonal basis memory requirement for $m = 35$ TT-vectors of order $d = 6$ and mode size $n = 15$. The curves get dashed and partially transparent when their corresponding loss of orthogonality gets greater then the prescribed rounding accuracy $\delta$.} 
		\label{fig:mem:6d}
	\end{figure}

	After describing the TT-Householder algorithm's memory requirements, we compare the four orthogonalization schemes from a memory consumption perspective. To make a fair comparison, we consider both memory consumption perspective and loss of orthogonality, which is studied in Section~\ref{sec3:s1}. Figure~\ref{fig:mem:6d} displays the maximum TT-rank, the compression ratio and the gain for the orthogonal TT-vectors $\ten{q}_k$ generated by the orthogonalization schemes plus the Householder TT-vectors $\ten{u}_k$, for different accuracies $\delta$. It is important to note that $\ten{q}_k$ are not computed in many applications. The curves in the corresponding figure become dashed and partially transparent for every rounding accuracy $\delta$ when the corresponding loss of orthogonality exceeds $10^{-1}$. In all Figures~\ref{fig:mem:6d:rk:3},~\ref{fig:mem:6d:rk:5}, and~\ref{fig:mem:6d:rk:8} show that the maximum TT-rank of the orthogonal TT-vectors computed by TT-CGS and TT-Gram schemes stagnates around $10$. However, there is no clear theoretical justification for this phenomenon. The maximal TT-rank of $\ten{q}_k$ from the TT-Gram algorithm is theoretically bounded by $k$ multiplied by the maximal TT-rank of $\ten{a}_j$ for $j\in\{1, \dots, k\}$. When generating $\ten{a}_j$, they are rounded with a maximal TT-rank equal to $1$. In our experimental framework, the maximal TT-rank of $\ten{q}_k$ is bounded by $k$. Conversely, the maximal TT-rank of $\ten{q}_k$ from TT-CGS is bounded by $1 + k(k-1)/2$ knowing that the maximal TT-rank of $\ten{a}_i$ is bounded by $1$ for $i\in\{1, \dots, m\}$ in our experiments. In terms of the maximal TT-rank, TT-Gram outperforms TT-MGS and TT-Householder for basis sizes greater than $10$. However, TT-CGS sets a lower bound for the maximal TT-rank. It is important to note that the loss of orthogonality of TT-CGS and TT-Gram becomes greater than $10^{-1}$ around $k = 10$. On the other hand, the TT-MGS maximal TT-rank curve becomes dashed around $k = 20$, while the Householder curve never does. This means that the loss of orthogonality arrives much later for TT-MGS or may not arrive at all for TT-Householder. In Figure~\ref{fig:mem:6d:rk:3}, the maximal TT-rank of $\ten{q}_k$ from TT-MGS exceeds the maximal TT-rank of the Householder TT-vector $\ten{u}_k$ and the Householder orthogonal TT-vector $\ten{q}_k$ when the basis size $k$ reaches $20$ and $25$, respectively. Figure~\ref{fig:mem:6d:rk:5} shows a comparable relationship between the maximal TT-rank for Householder and for MGS generated orthogonal TT-vectors. However, the turning point occurs at different basis sizes: $25$ for the Householder TT-vector $\ten{u}_k$ and $30$ for $\ten{q}_k$. For the last rounding accuracy value $\delta = 10^{-8}$, the maximal TT-rank of the MGS orthogonal TT-vector reaches the Householder $\ten{q}_k$ when the basis size exceeds $15$, surpassing the maximal TT-rank of the Householder $\ten{u}_k$ at approximately the same basis size. These results are displayed in Figure~\ref{fig:mem:6d:rk:8}.
	
	\par In analyzing the memory requirements of the TT-Householder algorithm, we also examine the compression ratio of the TT-vectors that form the orthogonal basis generated by the four orthogonalization methods. The compression ratio curves in Figures~\ref{fig:mem:6d:cr:3}, ~\ref{fig:mem:6d:cr:5} and~\ref{fig:mem:6d:cr:8} have the same slopes as their corresponding maximal TT-rank curves, but they clearly demonstrate the memory needs. The orthogonal TT-vectors obtained from the TT-CGS and TT-Gram schemes require only about $1\%$ of the memory needed to store the full format tensors, as shown in Figure~\ref{fig:mem:6d:cr:3}, ~\ref{fig:mem:6d:cr:5}, and~\ref{fig:mem:6d:cr:8}. However, when the basis size exceeds $10$ and the TT-vectors become more collinear, the resulting basis from these schemes become very poor in terms of orthogonality. Both Figures~\ref{fig:mem:6d:cr:3} and~\ref{fig:mem:6d:cr:5} indicate that storing the basis TT-vectors generated by the TT-Householder scheme requires approximately $20\%$ of the memory needed to store those tensors in full format. Similarly only $10\%$ of the entire memory required for full format storage is necessary to store the Householder TT-vectors $\ten{u}_k$. Finally, for $\delta = 10^{-8}$, the cost of storing the Householder basis TT-vectors $\ten{q}_k$ is the same as storing them in full format, as shown in Figure~\ref{fig:mem:6d:cr:8}. However, based on the same figure, it is evident that storing the Householder TT-vectors, even for $\delta = 10^{-8}$, requires only about $30\%$ of the memory needed to store the same tensors in full format. This feature makes the TT-Householder algorithm highly appealing, as it is usually adequate to store only the Householder TT-vectors. The TT-Householder algorithm becomes even more advantageous when compared to the compression ratio curves of the TT-MGS, TT-MGS2 and TT-CGS2 algorithms. For all rounding accuracy values, the compression ratio curve of the TT-MGS and TT-MGS2 always reaches $1$, as shown in Figures~\ref{fig:mem:6d:cr:3}, ~\ref{fig:mem:6d:cr:5}, and~\ref{fig:mem:6d:cr:8}. This implies that the memory required by the TT-vectors from these schemes is the same as the memory needed to store the orthogonal basis tensors in full format. This consideration also applies for the compression ratio of the orthogonal basis generated by the TT-CGS2 for $\delta\in\{10^{-5}, 10^{-8}\}$. However, for $\delta = 10^{-3}$, the TT-vectors from TT-CGS2 only require $20\%$ of the memory needed to store the same tensors in dense format. Figures~\ref{fig:mem:6d:g:3},~\ref{fig:mem:6d:g:5}, and~\ref{fig:mem:6d:g:8} show the compression gain curves for the different rounding accuracy values $\delta$ with a similar behavior. The compression gain of TT-Gram has a peak at the beginning and then stabilizes around $10$ starting from $k = 10$. This means that during compression, the $j$-th basis TT-vector reduces the memory requirement by $10$ times for $j\ge k$. The gain curves of TT-MGS, TT-MGS2 and TT-CGS2 have a similar shape. They increase up to $k = 15$, then drop for the next $5$ iterations before rising again around $k = 22$. The gain curve of TT-Householder basis follows a similar pattern, growing to a peak before decreasing and then rising again during the last iterations. As previously observed, the Householder TT-vector gain curve slightly rises at the beginning, then it drops down and stagnates at a very low value from $k > 10$. Finally, the TT-CGS gain curve increases as the dimension $k$ of the basis increases. When the last basis TT-vector is rounded, only $0.001$ of the memory used to store the TT-vector before rounding is required. However, as indicated by the dash style, both the TT-CGS and TT-Gram bases have almost completely lost the orthogonality already at $k = 10$. 
	
	\subsection{Summary}
	Table~\ref{tab:orth} summarizes the computational costs for the tensor and matrix cases. 
	\begin{table}[h]
		\centering
		\scalebox{0.9}{
			\begin{tabular}{ll ccc cc ccc}
				\toprule
				&& \multicolumn{3}{c}{\textbf{Matrix}} && \multicolumn{3}{c}{\textbf{TT-vectors}} \\
				\cmidrule(lr){3-5} \cmidrule(lr){7-9}
				\textit{\small Algorithm}&& \thead{\textit{Computational}\\ \textit{cost in fp operations}} && $\norm{\I_k - Q_k^\top Q_k}$&& \thead{\textit{Computational}\\ \textit{cost in \texttt{TT-round}}} &&$\norm{\I_k - Q_k^\top Q_k}$\\
				\cmidrule(lr){3-5} \cmidrule(lr){7-9}
				\textbf{Gram} 			&& $\bigO(2nm^2)$ &&	$\bigO(u\kappa^2(A_k))$	&& $m$ &&	$\bigO(\delta\kappa^2(A_k))$	\\
				\textbf{CGS} 			&& $\bigO(2nm^2)$ &&	$\bigO(u\kappa^2(A_k))$	&& $m$ &&	$\bigO(\delta\kappa^2(A_k))$	\\
				\textbf{MGS}			&& $\bigO(2nm^2)$ &&	$\bigO(u\kappa(A_k))$	&& $m$ &&	$\bigO(\delta\kappa^2(A_k))$	\\
				\textbf{CGS2}			&& $\bigO(4nm^2)$ &&	$\bigO(u)$	&& $2m$ &&	$\bigO(\delta)$	\\	
				\textbf{MGS2}			&& $\bigO(4nm^2)$ &&	$\bigO(u)$	&& $2m$ &&	$\bigO(\delta)$	\\
				\textbf{Householder} 	&& $\bigO(2nm^2 - 2m^3/3)$ &&	$\bigO(u)$	&& $4m$ &&	$\bigO(\delta)$	\\
				\bottomrule
		\end{tabular}}
		\caption{Computational costs in floating point operations and in \texttt{TT-round} operations, and bounds for the loss of orthogonality, theoretical with respect to the unit round-off $u$ and conjectured ones with respect to the rounding accuracy $\delta$, for an input set of $m$ vectors and TT-vectors respectively.}
		\label{tab:orth}
	\end{table}
	In terms of memory footprint, the TT-Householder orthogonalization scheme, along with TT-CGS2 and TT-MGS2, is often the best option due to its stability property and the option to store only the TT-vectors $\ten{u}_i$. Figures~\ref{fig:mem:6d:cr:3},~\ref{fig:mem:6d:cr:5}, and~\ref{fig:mem:6d:cr:8} demonstrate that when the input TT-vectors are not highly collinear ($k < 15$), TT-MGS2 and TT-CGS2 achieve a compression ratio similar to that of TT-Householder. For a basis size between $15$ and $25$ (or $20$ for $\delta = 10^{-8}$), TT-Householder is more memory-expensive than TT-MGS, TT-MGS2 and TT-CGS2. Finally, as the input TT-vectors become more linearly dependent, the memory requirements of the TT-MGS2 and TT-CGS2 bases become greater or equal to those of both the TT-Householder basis and Householder TT-vector. However, in terms of orthogonality preservation, TT-MGS2 outperforms TT-Householder for every rounding accuracy, while TT-CGS2 outperforms TT-Householder for $\delta\in\{10^{-5}, 10^{-8}\}$. In terms of computationa cost, both TT-CGS2 and TT-MGS2 are cheaper than TT-Householder, requiring only $2m$ \texttt{TT-round} instead of $4m$.
	
	\section{Concluding remarks}\label{sec4}
	In the framework where the data representation accuracy is decoupled from computational accuracy, as previously proposed in~\cite{Agullo2022, Coulaud2022a,Iannacito2022}, we investigate the loss of orthogonality of six orthogonalization kernels in the tensor format. The Tensor Train~\cite{Oseledets2009} is the compressed format used to represent tensors. The orthogonalization methods considered are Classical and Modified Gram-Schmidt (CGS, MGS), their versions with re-orthogonalization (CGS2, MGS2), the Gram approach and the Householder transformation. Section~\ref{sec2} describes the generalization of these kernels to the tensor space in TT-format, relying on the compression function called \texttt{TT-round}. Section~\ref{sec3} presents the numerical experiments related to the loss of orthogonality and memory requirement of these kernels in TT-format. 
	\par As in the matrix case, the choice of the orthogonalization scheme among TT-Householder, TT-CGS2 and TT-MGS2 depends strongly on the purpose and on the available computing resources. TT-Householder requires less memory, but it is computationally more expensive and its orthogonality stagnates around the rounding accuracy. On the other hand, TT-MGS2 produces a basis of better orthogonality quality, as long as the input TT-vectors are not too collinear, and it is computationally cheaper than TT-Householder. The same considerations hold also for TT-CGS2, under the same hypothesis. The theoretical validation of the experimental results remains an open question and will be the focus of future research.